\documentclass[12pt,preprint]{aastex}

\usepackage{epsfig}

\def\Term#1 #2 #3/{\mbox{$\,^{#1}\!#2_{#3}$ }}
\def\Termo#1 #2 #3/{\mbox{$\,^{#1}\!#2^o_{#3}$ }}
\def\sterm #1 #2 #3/{\mbox{$\,_{#3}\!^{#1}\!#2$}}

\newcommand{\teff}{$T_{\rm eff}$} 

\begin{document}

\title{{\it Spitzer} Observations of M33 and the Hot Star, \ion{H}{2} Region Connection}

\author{Robert H. Rubin\altaffilmark{1,2},
Janet P. Simpson\altaffilmark{1,3},
Sean W.J. Colgan\altaffilmark{1},
Reginald J. Dufour\altaffilmark{4}, 
Gregory Brunner\altaffilmark{4}, 
Ian A. McNabb\altaffilmark{1},
Adalbert W.A. Pauldrach\altaffilmark{5},
Edwin F. Erickson\altaffilmark{1}, 
Michael R. Haas\altaffilmark{1}, 
and
Robert I. Citron\altaffilmark{1}} 

\email{rubin@cygnus.arc.nasa.gov}

\altaffiltext{1} {NASA/Ames Research Center, Moffett Field, CA
94035-1000, USA}
\altaffiltext{2} {Orion Enterprises, M.S. 245-6, Moffett Field, CA
94035-1000, USA}
\altaffiltext{3} {SETI Institute, 515 N. Whisman Road, Mountain View, CA
94043, USA}
\altaffiltext{4}{Physics \& Astronomy Department, Rice University, MS 61, 
Houston, TX 77005-1892, USA}
\altaffiltext{5}{University of Munich,  Munich D-81679, Germany}

\def\Termo#1 #2 #3/{\mbox{$\,^{#1}\!#2^o_{#3}$ }}

\date{\today}

\begin{abstract}
We have observed emission lines of 
[\ion{S}{4}] 10.51,
H(7--6) 12.37,
[\ion{Ne}{2}] 12.81,
[\ion{Ne}{3}] 15.56,
and [\ion{S}{3}] 18.71~$\mu$m
in  a number of extragalactic \ion{H}{2} regions with the 
{\it Spitzer Space Telescope}. 
A previous paper presented our data and analysis for the substantially 
face-on spiral galaxy M83.
Here we report our results for the local group spiral galaxy M33.
The nebulae selected cover a wide range of galactocentric radii (R$_G$).  
The observations were made with the Infrared Spectrograph with the 
short wavelength, high resolution module.
The above set of five lines is observed cospatially, 
thus permitting a reliable comparison of the fluxes.
 From the measured fluxes,
we determine the ionic abundance ratios including
Ne$^{++}$/Ne$^+$,
S$^{3+}$/S$^{++}$, and S$^{++}$/Ne$^+$ 
and find that there is a correlation of increasingly higher ionization with 
larger R$_G$.
By sampling the dominant ionization states of Ne (Ne$^+$, Ne$^{++}$)
and S (S$^{++}$,   S$^{3+}$)
for \ion{H}{2} regions,  we can estimate the Ne/H, S/H, and Ne/S ratios.
We  find from linear least-squares fits that there is a decrease in metallicity
with increasing R$_G$:
d~log~(Ne/H)/dR$_G$~= $-0.058\pm0.014$ and
d~log~(S/H)/dR$_G$~= $-0.052\pm0.021$~dex~kpc$^{-1}$.
There is no apparent variation in the Ne/S ratio with R$_G$.
Unlike our previous similar study of M83, where we conjectured that
this ratio was an upper limit, for M33
the derived ratios are likely a robust indication of Ne/S.
This occurs because the \ion{H}{2} regions have lower metallicity and 
higher ionization than those in M83.
Both Ne and S are primary elements produced in $\alpha$-chain reactions, 
following C and O burning in stars,
making their yields depend very little on the stellar metallicity. 
Thus, it is expected that Ne/S remains relatively constant throughout a galaxy.
The median (average) Ne/S ratio derived for 
\ion{H}{2} regions in M33 is 16.3 (16.9), 
just slightly higher than the Orion Nebula 
value of 14.3.
The same methodology is applied to {\it Spitzer} observations
recently published for three massive \ion{H}{2} regions:
NGC~3603 (Milky Way), 30~Dor (LMC), and N~66 (SMC)
as well as for a group of blue compact dwarf galaxies.
We find median Ne/S values of 14.6, 11.4, 10.1, and 14.0, respectively.
All of  these values are in sharp contrast with the much lower 
``canonical", but controversial, solar value of $\sim$5.
A recent nucleosynthesis, galactic chemical evolution (GCE) model
predicts a Ne/S abundance of  $\sim$9.
	Our observations may also be used to test 
the predicted ionizing spectral energy distribution of various 
stellar atmosphere models.
We compare the ratio of fractional ionizations
$<$Ne$^{++}$$>$/$<$S$^{++}$$>$, 
$<$Ne$^{++}$$>$/$<$S$^{3+}$$>$, and
$<$Ne$^{++}$$>$/$<$Ne$^+$$>$
vs.\ 
$<$S$^{3+}$$>$/$<$S$^{++}$$>$
with predictions
made from our photoionization models using 
several of the state-of-the-art stellar atmosphere model grids.
The trends of the ionic ratios established from the prior M83 study
are remarkably similar, but continued to higher ionization with the 
present M33 objects.
\end{abstract}

\keywords{ISM: abundances, \ion{H}{2} regions, stars: 
atmospheres, galaxies: individual (M33)}

\vspace{12pt}

\section{Introduction}

	This work is a continuation of a similar previous study
of 24 \ion{H}{2} regions in the substantially face-on spiral galaxy M83
that we observed with the {\it Spitzer Space Telescope} 
(Rubin et~al.\ 2007, hereafter R07).
Here we present our more recent {\it Spitzer} observations of
25 \ion{H}{2} regions in the local group spiral galaxy M33.
The analyses of the new set, in conjunction with the M83 set,
further elucidate topics investigated in  R07.
Since this paper should stand alone, that is, not require the reader 
to fully read R07 first, much of the material presented there
is repeated here.
At times, this may be verbatim.

      Most observational studies of the chemical evolution of the universe
rest on emission line objects, which define the mix of elemental abundances at
advanced stages of evolution as well as 
the current state of the interstellar medium (ISM).
Gaseous nebulae 
are laboratories for understanding physical 
processes in all emission-line sources and probes for stellar, 
galactic, and primordial nucleosynthesis. 
{\it Spitzer} has a unique ability to address the
abundances of the elements neon and sulfur.
This is particularly true in the case of \ion{H}{2} regions,
where one can observe simultaneously four emission lines that
probe the dominant ionization states of Ne (Ne$^+$ and Ne$^{++}$)
and S (S$^{++}$ and S$^{3+}$).
The four lines, 
[\ion{Ne}{2}] 12.81,
[\ion{Ne}{3}] 15.56,
[\ion{S}{3}] 18.71, and [\ion{S}{4}] 10.51~$\mu$m
can be observed cospatially 
with the Infrared Spectrograph (IRS) on the {\it SST}.
Because of the sensitivity of {\it SST},
the special niche, relative to previous (and near-term foreseeable)
instruments, is for studies of  extragalactic \ion{H}{2} regions.
Toward this end, we have used {\it SST} to 
observe $\sim$25 \ion{H}{2} regions each in galaxies with 
various metallicities and other properties.
To the extent that all the major forms of Ne and S
are observed, the true Ne/S abundance ratio could be inferred.
For Ne, this is a safe assumption, but for S, there is the possibility 
of non-negligible contributions due to S$^+$ as well as what could be tied up in
molecules and dust.
Due to this, we surmised that the large values derived for Ne/S for the M83 
\ion{H}{2} regions, which are fairly low ionization, were in fact {\it upper limits}.
With a portion of our Cycle 2 {\it SST} observations of 
M33 \ion{H}{2} regions in hand, we concluded that the preliminary Ne/S results 
(from $\sim$12 to 21) were likely a reliable estimate because these nebulae had
lower metallicity and much higher ionization than those in M83
(R07, Rubin~et~al.\ 2006).

	The preliminary M33 Ne/S ratios and the Orion Nebula value of
14.3 (Simpson et~al.\ 2004) were in sharp contrast with the much lower 
``canonical" solar value ($\sim$5) 
(Lodders 2003;  Asplund, Grevesse,  \& Sauval 2005).
There was an even larger difference compared with the Ne/S
ratio predicted by GCE models. 
According to calculations based on
the theoretical  nucleosynthesis, galactic
chemical  evolution models  of Timmes, Woosley,  \& Weaver (1995),
the Ne/S ratio in the solar neighborhood would change little,
from 3.80 to  3.75, between solar birth and the present
time  (apropos  for the Orion Nebula).
These calculations were provided by Frank Timmes (private communication).
Timmes noted that although massive stars
are expected to dominate the Ne and S production (and are all that
were included in their non-rotating models with no wind losses),
there is likely to be some re-distribution of Ne and S from rotation 
or from Wolf-Rayet phases of evolution, along with contributions 
of Ne from novae or even heavier intermediate mass stars.
Hence, the Ne/S of 3.8 should be considered
a lower bound because some potential sources of Ne are missing.
With {\it SST} observations such as  we have undertaken for
M83  and M33, there is important feedback to the  GCE field. 

	The presence of radial (metal/H) abundance gradients in the plane 
of the Milky Way is well established in both gaseous nebulae and stars
(e.g., Henry \& Worthey 1999; Rolleston~et~al.\ 2000).
Radial abundance gradients seem to be ubiquitous in spiral galaxies, 
though the degree varies
depending on a given spiral's morphology and luminosity class. 
The gradients are 
generally attributed to the radial dependence of star formation 
history and ISM mixing processes (e.g., Shields 2002).
Thus, the observed gradients are another tool for understanding galactic evolution 
(e.g., Hou~et~al.\ 2000; Chiappini et~al.\ 2001; Chiappini et~al.\ 2003).
The premise is that star formation and chemical enrichment begins in the 
nuclear bulges of the galaxies and subsequently progresses outward into 
the disk, which has remained gas-rich.
The higher molecular gas density in the inner regions produces a higher star 
formation rate, which results in a 
relatively greater return to the ISM of both ``primary'' 
$\alpha$-elements (including O, Ne, and S)
from massive star supernovae, and ``secondary'' elements like N. 
Secondary nitrogen is produced by CNO burning of already existing carbon and oxygen 
in intermediate-mass stars and is subsequently returned to the ISM through mass loss.
However, because chemical evolution models have
uncertain input parameters, and because
details of the abundance variations of each element
are uncertain,
current understanding of the formation and evolution of galaxies suffers
(e.g., Pagel 2001).

	Studies of \ion{H}{2} regions
in the Milky Way are hampered by interstellar extinction.
For the most part, optical studies (e.g., Shaver~et~al.\  1983) 
have been limited to those \ion{H}{2} regions
at galactocentric radius R$_G \gtrsim 6$ kpc
(predicated on $R_\odot = 8$ kpc) 
because \ion{H}{2} regions are very concentrated to the Galactic plane.
Here extinction becomes severe with increasing distance from Earth.
Observations using far-infrared (FIR) emission lines
have penetrated the R$_G \lesssim 6$~kpc barrier. 
Surveys with the {\it Kuiper Airborne Observatory (KAO)}
by Simpson~et~al.\ (1995),
Afflerbach~et~al.\ (1997),
and Rudolph~et~al.\ (2006)
have observed 16 inner Galaxy \ion{H}{2} regions.
With the {\it Infrared Space Observatory (ISO)},
Mart\'{\i}n-Hern\'{a}ndez~et~al.\ (2002a) observed 13 inner Galaxy 
\ion{H}{2} regions covering FIR and also mid-IR lines.
A major finding of these studies is that inner Galaxy \ion{H}{2} 
regions generally have lower excitation (ionization)
compared to those at larger R$_G$.
This holds for both heavy element ionic ratios O$^{++}$/S$^{++}$ 
(Simpson~et~al.\ 1995) and 
Ne$^{++}$/Ne$^{+}$ (Simpson \& Rubin 1990; Giveon~et~al.\ 2002),
and also He$^+$/H$^+$ measured from radio recombination lines 
(Churchwell~et~al.\ 1978; Thum, Mezger, \& Pankonin 1980).
Whether the observed increase in  excitation with increasing R$_G$ 
comes  entirely from heavy element opacity effects 
in the \ion{H}{2} regions and stellar atmospheres,
or also from 
a gradient in the maximum stellar effective temperature, \teff, 
of the exciting stars 
is still a point of controversy
(e.g., Giveon~et~al.\ 2002; 
Mart\'{\i}n-Hern\'{a}ndez~et~al.\ 2002b; 
Smith, Norris, \& Crowther 2002; Morisset~et~al.\ 2004).

	It has become clear that nebular plasma simulations with 
photoionization modeling codes are enormously sensitive
to the ionizing spectral energy distribution (SED) that is input
(e.g., R07 and Simpson et~al.\ 2004, and references in each).
These SEDs need to come from stellar atmosphere models.
In R07, we developed new observational tests of and constraints 
on the ionizing SEDs that are predicted from various stellar atmosphere models.
We compared our {\it SST} observations of
\ion{H}{2} regions in M83 with the various tracks predicted from
photoionization models that changed {\it only the ionizing SEDs}.
M83 provided data for high metallicity 
(at least twice solar, e.g., Dufour et~al.\ 1980;
Bresolin \& Kennicutt 2002) and lower-ionization \ion{H}{2} regions.
The best overall fit to the nebular models was obtained using the
supergiant stellar atmosphere models  
computed with the WM-BASIC code (Pauldrach, Hoffmann, \& Lennon 2001;
Sternberg, Hoffmann, \& Pauldrach 2003).

	We discuss the M33 {\it SST}/IRS observations in section 2. 
In section 3, the data are used to test for
a variation in the degree of ionization of the
\ion{H}{2} regions with R$_G$.
We examine the Ne/S abundance ratio for our 
M33 \ion{H}{2} region sample in section 4.
In section 5, we test for 
a variation in the Ne/H and S/H ratio with R$_G$.
Section 6 describes how these {\it Spitzer} data are used to constrain
and test the ionizing SEDs predicted by stellar atmosphere models.
In section 7, there is additional discussion pertaining to the
Ne/S ratio.
Last, we provide a summary and conclusions in section~8.

\section{{\it Spitzer Space Telescope} Observations}

	In the substantially face-on (tilt 56$^{\rm o}$) local group,
spiral galaxy M33,
we observed 25 \ion{H}{2} regions, 
covering a wide range of deprojected galactocentric radii (R$_G$)
from 0.71 to 6.73~kpc.
We used the {\it SST}/IRS in 
the short wavelength, high dispersion (spectral resolution $\sim$ 600) 
configuration,
called the short-high (SH) module
(e.g.,  Houck~et~al.\ 2004).
This covers the wavelength range from 9.9~-- 19.6~$\mu$m
permitting cospatial observations of all five of our programme 
emission lines:
[\ion{S}{4}] 10.51,
hydrogen H(7--6) (Hu$\alpha$)  12.37,
[\ion{Ne}{2}] 12.81,
[\ion{Ne}{3}] 15.56,
and [\ion{S}{3}] 18.71~\micron.

	The data were collected under the auspices of Spitzer
programme identification 20057.
Most of  the observations were made in 2006, January 15 to 
February 1 (UT) during {\it SST}/IRS campaign 28.
The last set of observations, one Astronomical Observing Request  (AOR)
out of a total of 8,  
was made on 2007, February 11 during {\it SST}/IRS campaign 38.
Figure~1 shows the regions and apertures observed,
while Table~1 lists the \ion{H}{2} region positions
and the aperture grid configuration used to observe each.
The nebulae are designated by their BCLMP number  
(Boulesteix et~al.\ 1974).
The \ion{H}{2} regions observed in the remaining AOR
in 2007  were \#62, 302, 691, 651, and 740W.
The size  of the SH aperture is 11.3$''$$\times$4.7$''$. 
Maps were arranged with the apertures overlapping  
along the direction of the long  slit  axis
(the ``parallel" direction).
The purpose of overlapping is that most spatial positions will be 
covered in at least two locations on the array, minimizing the effects 
of bad detectors.
In the direction of  the short slit axis 
(the ``perpendicular" direction),
the apertures were arranged immediately
abutting each other; that is, with no overlap or space between them.
In all cases, we chose the mapping mode  with 
aperture grid patterns varying from a
1$\times$2 grid to as  large as a 2$\times$4 grid
in order to cover the bulk of the expected emission.

	Our terminology for  the aperture grid pattern is intended to
give roughly the map size in integer multiples of the parallel $\times$
perpendicular direction aperture size.
For instance the 1$\times$2 and 1$\times$3  patterns
have a shift of 3.45$''$ in the parallel direction
and a 4.7$''$ shift in the perpendicular direction.
The resulting map has a full size in the 
parallel direction of 14.75$''$ (11.3 + 3.45) which is  closest to 
the integer one in the parallel direction.
For the two other aperture grid patterns used,
the 2$\times$3 and 2$\times$4  patterns,
the shift (or step)  is 3.77$''$ (1/3 the aperture length)
in the parallel direction.
Thus, the resulting maps have a full size that covers an
area of 2$\times$3 or 2$\times$4  apertures, respectively.
In order to save overhead time, we clustered the objects into 
AORs with the same aperture grid pattern.
We used a ramp (exposure) time of 30~s  and 12 cycles  at each
spatial position.
This permits up to effectively 720~s (24 cycles)
integration time for some spatial positions in the
1$\times$2 and 1$\times$3  patterns and up to 1080~s (36 cycles) 
for some positions in the 2$\times$3 and 2$\times$4 patterns.

	Our data were processed and calibrated 
with version S15.3.0 of the standard IRS pipeline at the 
{\it  Spitzer} Science Center.
We use CUBISM,
the CUbe Builder for IRS Spectral Mapping, (version 1.50) to build 
our post-BCD
(basic  calibrated data)
 data products.  
CUBISM is described in  Smith  et~al.\ (2007a) and references therein,
as well as in a manual detailing its use (Smith et~al.\ 2007b).
CUBISM was used to build maps, including accounting for aperture overlaps,
and to deal effectively with bad pixels.  
   From the IRS mapping observations, it can combine these data
into a single 3-dimensional cube with two spatial and one spectral dimension.
For each of our programme \ion{H}{2} regions,
we constructed a data cube.
Global bad pixels (those occurring at the same pixel in every BCD) were
removed manually.  
Record level bad pixels (those occurring only within individual BCDs) 
that deviated by 5~$\sigma$ from the median pixel value
and occurred within at least 10~per~cent of the BCDs were removed 
automatically in CUBISM with the ``Auto Bad Pixels" function.  
In reducing our data, we were careful to monitor that the 
``Auto Bad Pixels" function did not incorrectly flag any of the pixels 
on our programme spectral lines as bad.
Data cubes for each \ion{H}{2} region were built without applying the 
slit-loss correction factors (SLCFs).  
This is discussed below.
We varied our spatial extraction aperture size [always a 2-D integer pixel
grid] due to the differences in the size of the observing grid over the 
particular nebula.
Our further analysis of these spectra uses
the line-fitting routines in  the IRS Spectroscopy Modeling Analysis 
and Reduction Tool (SMART, Higdon et~al.\ 2004).

The emission lines were measured with SMART using a Gaussian line fit.
The continuum baseline was fit with a linear or
quadratic function.
Figures~2 (a)--(e) show the fits for each of the five lines
in BCLMP~45 (object \#6 in Fig.~1).
A line is deemed to be detected if the flux is  at least
as  large as the 3~$\sigma$ uncertainty.
We measure the uncertainty by the product of the full-width-half-maximum
(FWHM) and the
root-mean-square variations in the adjacent, line-free continuum;
it does not include systematic  effects.

	In section 2 of our M83 paper (R07), 
we discussed and estimated systematic uncertainties
as they affected the line fluxes. 
Here we recap (or repeat) the major points.
Most likely the largest uncertainty is due to slit (aperture) loss 
correction factors (SLCFs).
The pipeline flux calibration assumes that objects are point sources.
Our nebulae are extended and that is why  we mapped each with a grid  that
covers more than a single aperture.  
We did not make a correction for this effect.
Thus we have implicitly assumed that the \ion{H}{2} regions are 
close to the point-source limit within the SH 11.3$''$$\times$4.7$''$ aperture.
If the \ion{H}{2} region were uniformly extended within the SH aperture,
SLCFs would need  to  be applied  to our  fluxes.
These are: 0.697, 0.671, 0.663, 0.601, and 0.543 for the 
10.5, 12.4, 12.8, 15.6, and 18.7~$\mu$m lines, respectively.
These factors were obtained by interpolating in numbers provided from 
the $`$$b1\_slitloss\_convert.tbl$$'$ file from the
{\it Spitzer} IRS Custom Extraction tool (SPICE) for the SH module.
For the uniformly filled aperture,  the 
maximum uncertainty in the flux due to this effect
would be $\sim$46~per~cent for the [\ion{S}{3}] 18.7 line.  
The SLCFs would need to multiply our listed fluxes.
We note that with regard to this effect, the fluxes listed in Table~2
are  upper limits and that the uncertainty would be only in the
direction to lower them.
No correction factor was applied because we are likely closer to
the point-source limit than the uniform-brightness limit.
Because our science depends on line flux ratios, for our purposes, the 
possible  uncertainty due to this effect would be lower, e.g.,
$\sim$22~per~cent when we deal with the line flux ratio 
[\ion{S}{4}] 10.5/[\ion{S}{3}] 18.7.
The possible uncertainty in the absolute
flux calibration of the spectroscopic products delivered by the pipeline
is likely confined to between $\pm$5~per~cent  and $\pm$10~per~cent.  
	Any uncertainty in the flux due to a pointing error is probably small
and in the worst case should not exceed 10~per~cent.  

	For the brighter lines, that is, most of the
10.5, 12.8, 15.6, and 18.7~$\mu$m lines,
the systematic uncertainty far exceeds the measured (statistical) uncertainty.
Even for the fainter lines, we estimate that the systematic uncertainty
exceeds the measured uncertainty.
In addition to the line flux, the measured FWHM and
heliocentric radial velocities (V$_{helio}$)
are listed in Table~2.
Both the FWHM and V$_{helio}$ are useful in judging the
reliability of the line measurements. 
The FWHM is expected to be the instrumental width for all our lines.  
With a resolving power for the SH module of $\sim$600, our lines should have a
FWHM of roughly 500~km~s$^{-1}$.
The values for 
V$_{helio}$ 
should straddle the heliocentric systemic radial
velocity for M33 of $-179$~km~s$^{-1}$
(Corbelli \& Schneider 1997).
Most of our measurements are in agreement with these expectations.

\section{Variation in the degree of ionization of the
\ion{H}{2} regions with R$_G$}

	We chose our sample of nebulae 
in order to cover a wide range in R$_G$ (in the plane of M33).  
To derive these deprojected galactocentric distances,
we used a distance D of 840~kpc 
(Freedman, Wilson, \& Madore 1991)
an inclination angle ($i$~= 56$\pm1^{\rm o}$),
and a position angle of the line of nodes ($\theta = 23\pm1^{\rm o}$)
(Zaritsky, Elston, \& Hill 1989).
We assumed the centre of the galaxy is at
$\alpha$, $\delta$ = $1^{\rm h}33^{\rm m}51\fs02$, 
$30^{\rm o}$39\arcmin36\farcs7 (J2000) 
(Cotton, Condon, \&  Arbizzani 1999).
Table~3 lists R$_G$ for the centre of each object.
These range from from 0.71 to 6.73~kpc.

     From the measured fluxes, we estimate ionic 
abundance ratios, including  Ne$^{++}$/Ne$^+$,
S$^{3+}$/S$^{++}$, and S$^{++}$/Ne$^+$, 
for each of the \ion{H}{2} regions.
Important advantages compared with prior optical studies 
of various other ionic ratios are: 
(1) the IR lines have a weak and similar electron temperature ($T_e$) 
dependence while the collisionally-excited
optical lines vary exponentially with $T_e$, and 
(2) the IR lines suffer far less from interstellar extinction.  
Indeed for our purposes,  the differential extinction correction
is negligible as the lines are relatively close in wavelength.
In our analysis, we deal with ionic abundance ratios
and therefore line flux ratios.
In order to derive the 
ionic abundance ratios, we perform the usual semiempirical
analysis assuming a constant $T_e$ and electron density ($N_e$)
to obtain the volume emissivities for the five pertinent transitions.
For the ions Ne$^+$, Ne$^{++}$, S$^{++}$, and S$^{3+}$,
we use the  atomic data described in Simpson et~al.\ (2004) and 
Simpson et~al.\ (2007). 
For H$^+$ from the H(7--6) line, we use Storey \& Hummer (1995). 
There is a bit of a complication here because at Spitzer's spectral 
resolution, the H(7-6) line is blended with the H(11-8) line.
Their respective $\lambda$(vac)~= 12.371898 and 12.387168~$\mu$m.
In order to correct for the contribution of
the H(11-8) line, we use the relative intensity of H(11-8)/H(7-6)
from recombination theory (Storey \& Hummer 1995) assuming
case~B and $N_e$~= 100~cm$^{-3}$.
The  ratio H(11-8)/H(7-6)~= 0.122 and holds over a fairly wide range
in $N_e$ and $T_e$ [including 8000 and 10000~K, see below]
appropriate for our objects, and indeed for case~A also.

	For the entries in Table~3, we adopt a value for all the M33 
\ion{H}{2} regions of $T_e$~= 8000~K
and $N_e$~= 100~cm$^{-3}$.
In a recent paper, Magrini et~al.\ (2007) were able to derive
$T_e$ for 14 \ion{H}{2} regions in M33 from the diagnostic flux ratio
[\ion{O}{3}]  4363/(5007~+~4959).
Only one of their objects BCLMP~45 can be clearly identified  with one
of our regions.
They found $T_e$[\ion{O}{3}]~= 8600$\pm$200~K.
Furthermore, this object was one of only three where they also
determined $T_e$[\ion{N}{2}]
from  the flux ratio  5755/(6584~+~6548);  the result was 8200$\pm$1000~K.
There are two more of our programme sources with available
$T_e$[\ion{O}{3}].
These were derived by Crockett et~al.\ (2006) from  their optical
observations of BCLMP~691 yielding 10000$\pm$200~K
and recomputing $T_e$ with the measurements for 
BCLMP~280 (NGC~588) from 
V\'ilchez et~al.\ (1988) 
resulting in 9300$^{+600}_{-400}$.
While these values for  $T_e$ are somewhat higher than the
8000~K we adopt, we point  out a well-known bias.
That is,  both $T_e$[\ion{O}{3}] and $T_e$[\ion{N}{2}] derived from
the ratio of fluxes of ``auroral" to ``nebular" lines
are systematically higher than the so-called ``$T_0$",
which is the ($N_e$$\times$$N_i$$\times$$T_e$)--weighted average, where  
$N_i$  is  the ion  density of  interest.
The amount of  this bias depends on the degree of $T_e$
variations in the observed volume
(see Peimbert 1967 and many forward references).
In our analysis,  using the set of  IR lines, it is more 
appropriate to be using a $T_e$ that is similar to $T_0$.
Because  of the insensitivity of the volume emissivities
to $T_e$, particularly when working with ratios for these IR lines,
our results depend very little on this $T_e$ choice. 
The effects on our analysis due to a change in the assumed
$N_e$ are also small as will be discussed later.

	We present the variation of Ne$^{++}$/Ne$^+$ with R$_G$
in Figure~3 using the values from Table~3.
The error values here, as well as for all others in
Table~3
and in Figures 4--8, represent the propagated flux
measurement uncertainties
and {\it do not include the systematic uncertainties}.
Our assumed $N_e$ of 100~cm$^{-3}$ appears reasonable in view
of previous observations that address  the density.
For instance, Magrini et~al.\ (2007) derived $N_e$  from the
familiar diagnostic line flux  ratio [\ion{S}{2}] 6717/6731.
For most of their  \ion{H}{2} regions, they found that this
ratio was consistent with the  low-$N_e$ asymptotic limit,
that is, $N_e$~$<$ 100~cm$^{-3}$.
There is extremely little change in any of our derived
Ne$^{++}$/Ne$^+$ ratios
even when using an $N_e$  of 1000~cm$^{-3}$, which is likely 
an upper limit for these \ion{H}{2} regions.
A linear least-squares fit indicates a positive
correlation with R$_G$ (in kpc),

~~~~~~~~~~~~~~~Ne$^{++}$/Ne$^+$~= $-$0.44$\pm$0.22~+~(0.46$\pm$0.067)~R$_G$,

\noindent
with miniscule change to this equation for $N_e$~= 1000~cm$^{-3}$.
For all the least-squares line fits in this paper,
each point is given equal weight
because systematic uncertainties exceed the flux measurement 
uncertainties, as discussed earlier.
The positive correlation of Ne$^{++}$/Ne$^+$  
with R$_G$ as measured by the slope may
be judged to be significant following the criterion that it
exceeds the 3~$\sigma$ uncertainty.
We also did a linear least-squares fit to  log(Ne$^{++}$/Ne$^+$) vs.\ R$_G$
with the result

~~~~~~~~~~~~~log(Ne$^{++}$/Ne$^+$)~= $-$0.88$\pm$0.11~+~(0.20$\pm$0.035)~R$_G$.

\noindent
This relation also produces a statistically significant slope.
The transformation of this function is shown as 
the dashed  line  in Figure~3.
Three objects (230, 702,  and 740W) have been excluded in this 
Figure as well as in Figure~4 because of poor S/N and/or extreme deviancy
from the trend of  the 22 other sources.
The slope in the linear plot here 0.46$\pm$0.067
is much steeper than the analogous slope for our M83 \ion{H}{2} regions
of 0.011$\pm$0.0035  (see  Figure~3  in R07).
Furthermore, the comparison with Figure~3 in the M83 paper
shows dramatically the higher ionization of the M33 nebulae.

	A similar fit to the 	
S$^{3+}$/S$^{++}$  vs.\ R$_G$ data yields

~~~~~~~~~~~~~~S$^{3+}$/S$^{++}$~= $-$0.032$\pm$0.018~+~(0.046$\pm$0.0057)~R$_G$.

\noindent
The slope exceeds  the 1~$\sigma$ uncertainty by a factor of 8.
Thus the increase in degree of ionization with increasing
R$_G$ here too is significant.
A linear least-squares fit to log(S$^{3+}$/S$^{++}$) vs.\ R$_G$
results in

~~~~~~~~~~~~~log(S$^{3+}$/S$^{++}$)~= $-$1.7$\pm$0.090~+~(0.18$\pm$0.028)~R$_G$.

\noindent
Again there is a statistically significant slope.
We map this relation onto  Figure~3 as the dashed line.
While there is no fundamental reason to  expect either functional
form shown in Figures 3  and  4, we note  that the dashed line fits
do have the advantage of not extrapolating to negative ionic ratios
at small values of R$_G$.

	Figure~5 plots the fractional ionic abundance ratio
$<$S$^{++}$$>$/$<$Ne$^+$$>$ vs.\ R$_G$ for 23 \ion{H}{2} regions
(sources 230 and 702 are excluded).
This ratio is obtained from the 
S$^{++}$/Ne$^+$ ratio by multiplying by an assumed Ne/S value (see below).
The last three columns of Table~3 list this and other
fractional ionic abundance ratios used in this paper.
We show the linear least-squares fit for an assumed $N_e$ of 100~cm$^{-3}$.
Here,  the fit indicates a significant positive correlation with R$_G$,

~~~~~~~~~~~~~~~$<$S$^{++}$$>$/$<$Ne$^+$$>$~= 0.56$\pm$0.23~+~(0.31$\pm$0.069)~R$_G$, 

\noindent
where angular brackets denote fractional ionization.
In this figure and in the linear fit, we assume an Orion Nebula Ne/S 
abundance ratio of 14.3 (Simpson et~al.\ 2004).
Because Ne and S are ``primary" elements,
their production is expected to vary in lockstep
and Ne/S would not be expected to show a radial gradient within a galaxy
(Pagel \& Edmunds 1981).
There is a clear correlation of increasingly higher ionization with 
increasing R$_G$.
One reason may be due to the lower metallicity at larger R$_G$  
(see section 5)
causing the exciting stars to have a harder ionizing spectrum.  

\section{Neon to Sulfur abundance ratio}

	For \ion{H}{2} regions, we may approximate the Ne/S ratio with
(Ne$^+$ + Ne$^{++}$)/(S$^{++}$ + S$^{3+}$).
This includes the dominant ionization states of these two elements.
However this relation does not account for S$^+$,
which should  be present at some level.
We may safely ignore the negligible contributions
of neutral Ne and S in the ionized region.
Figure~6 shows our approximation for Ne/S vs.\ R$_G$.
The linear least-squares fit to 23 objects (again omitting for cause
sources 230 and 702) is

~~~~~~~~~~~~~~~Ne/S~= 18.0$\pm$1.3~$-$~(0.39$\pm$0.40)~R$_G$,

\noindent
plotted as the dotted line in Figure~6.
We also show the fit, the solid line, after removing the remaining deviant large
value (source 32).
The relation becomes

~~~~~~~~~~~~~~~Ne/S~= 16.8$\pm$0.97~$-$~(0.14$\pm$0.29)~R$_G$.

\noindent
Both of these indicate that there is no significant slope.
Our data also  indicate that the  lower envelope to  Ne/S
is well fit by a constant value equal to the Orion Nebula ratio
of 14.3 (Simpson~et~al.\ 2004). 

	In our previous M83 results, there appeared to be a 
drop in the Ne/S ratio with increasing R$_G$.
We argued  that this was not a true gradient in Ne/S.
Instead, it is most likely due to not accounting for the
presence of sulfur in other forms~-- S$^+$, molecules, and dust.
Because our observations of the \ion{H}{2} regions in M83
showed an increasing degree of ionization  with increasing R$_G$,
the expected increasing fraction of S$^+$ towards the inner 
galaxy regions would lead to a flatter gradient.
Another factor that could flatten the slope is the higher dust content
(with S, but  not Ne, entering grains)
expected in the inner regions due to higher metallicity
as is the case for the Milky Way.
The refractory carbonaceous and silicate grains are not distributed
uniformly throughout the Galaxy but instead increase in density toward the
centre.
A simple  model suggests the dust density is $\sim$5~-- 35 times higher
in the inner parts  of the Galaxy than in the local ISM 
(Sandford,  Pendleton,  \& Allamandola 1995).
The Ne/S abundance ratios that we derived for 24 
\ion{H}{2} regions in M83 varied from 41.9 to 24.4
(see Figure~5 in R07).
All are considerably higher than the Orion Nebula value
of 14.3 and, as an ensemble, significantly
higher  than what we find here for the nebulae  in M33.
Thus the evidence is strong that the derived Ne/S estimates 
for the M83 objects are upper limits and furthermore,
those that are further from the centre will
likely need less of a downward correction to obtain a true Ne/S ratio.

Because the M33 \ion{H}{2} regions have a lower metallicity 
and because almost all have a significantly higher ionization than those 
we observed in M83, the amount of any correction needed for
S in forms other than S$^{++}$ and S$^{3+}$ is minimized.
Hence while our derived Ne/S ratios should still be considered
as upper  limits,  these M33 values are a much more robust
estimate of a true Ne/S ratio than those for M83.
The scenario above is in excellent accord with the recent results of
Wu~et~al.\ (2008), hereafter W08.
They  obtained an average Ne/S~= 12.5$\pm$3.1
from  {\it Spitzer} observations of  13 blue compact dwarf galaxies
and found no correlation  between their  Ne/S and Ne/H ratios.
The median (average) Ne/S ratio derived for 23 \ion{H}{2} regions in M33 is 
16.3 (16.9).

\section{Variation in the Ne/H and S/H ratios with R$_G$}

There was a significant detection of the H(7-6) flux 
for 16 of the \ion{H}{2} regions (see Table~2) which permits a 
determination of the heavy element abundances Ne/H and S/H (see Table~3).
We present the results for   Ne/H in Figure 7.  
A linear least-squares fit of log~(Ne/H) vs.\ R$_G$ results in

~~~~~~~~~~~~~~~log~(Ne/H)~= $-$4.07$\pm$0.04~$-$~(0.058$\pm$0.014)~R$_G$.

\noindent
This fit  indicates a significant slope (4.1~$\sigma$). 
A similar plot for S/H is shown in Figure 8.  
Here the linear least-squares fit to log~(S/H) vs. R$_G$ yields

~~~~~~~~~~~~~~~log~(S/H)~= $-$5.31$\pm$0.06~$-$~(0.052$\pm$0.021)~R$_G$.

\noindent
It is interesting that the slope is nearly identical to the Ne/H relationship.  
However, the slope signal-to-noise ratio
(2.5~$\sigma$) is less than 3~$\sigma$ 
and thus would be deemed only marginally significant.
While {\it  Spitzer} is an admirable machine for measuring both Ne and S
abundances in \ion{H}{2} regions, the neon abundances are determined
more reliably.
As previously mentioned, with the {\it Spitzer} observations alone,
we are neither accounting for S$^+$ nor  S  that  may  be
tied up  in dust or  molecules.
In this  sense,  the S/H ratios in Figure~8 are lower limits.
Both of  the above measurements  of a heavy element abundance
gradient  are in remarkable agreement  with  the recent
value  for the log~(O/H) gradient of  
$-$0.054$\pm$0.011~dex~kpc$^{-1}$ (Magrini~et~al.\ 2007).
They derived this gradient from  optical observations  of 14 \ion{H}{2} 
regions in  M33 where the [\ion{O}{3}] (5007, 4959, 4363~\AA) and 
[\ion{O}{2}] (7320, 7330~\AA) emission line  fluxes  were all measured 
and a value for $T_e$  determined.
The sources for their linear least-squares fit covered a  range
in R$_G$  from $\sim$2 to  7.2~kpc  and  resulted in
log~(O/H)~= $-$3.47$\pm$0.05~$-$~(0.054$\pm$0.011)~R$_G$.
Because the slope for this is practically the same as ours above
for  log~(Ne/H), we may use the respective y-intercepts to
infer a Ne/O ratio of  0.28.
This is in good agreement with the Ne/O~= 0.25 value 
for the Orion Nebula  (Simpson et~al.\ 2004).
Using {\it ISO}, Willner \& Nelson-Patel (2002) measured the 
neon lines in M33.
If the two outermost objects in their sample are neglected,
they found a neon gradient of $-$0.05$\pm$0.02~dex~kpc$^{-1}$,
which agrees with the slope  we  find.

\section{Constraints on the ionizing SED for the stars exciting the
\ion{H}{2} regions}

	Various fractional ionic abundances are highly sensitive
to the stellar ionizing SED that apply to \ion{H}{2} regions.
The present {\it Spitzer} data probe the 
Ne$^+$ and Ne$^{++}$ fractional ionic abundances,
as well as those of  S$^{++}$ and S$^{3+}$.
They may be used to provide further constraints  and tests 
on the ionizing SED for the stars exciting these M33 nebulae,
similar to what we had done in the M83 paper (R07).
We use the 
ratio of fractional ionizations
$<$Ne$^{++}$$>$/$<$S$^{++}$$>$ 
vs.\ 
$<$S$^{3+}$$>$/$<$S$^{++}$$>$
(Figure~9a), 
$<$Ne$^{++}$$>$/$<$S$^{3+}$$>$
vs.\ 
$<$S$^{3+}$$>$/$<$S$^{++}$$>$
(Figure~9b),
and Ne$^{++}$/Ne$^+$ vs.\
$<$S$^{3+}$$>$/$<$S$^{++}$$>$
(Figure~9c). 
These ionic ratios are computed 
using our photoionization code NEBULA
(e.g., Simpson et~al.\  2004; Rodr\'\i guez \& Rubin 2005).
The lines connect the results of the nebular models calculated using the 
ionizing SEDs predicted from various stellar atmosphere models.
There are \underbar{no other changes} to the input parameters,
just the SED.
The stellar atmospheres used are  representative of several
non-LTE models that apply for O-stars.
We also  display the results from one set of LTE models (Kurucz 1992).  
His LTE atmospheres have been extensively used in
the past as input for \ion{H}{2} region models.
Hence the comparison with the other non-LTE results reinforces the fact 
that more reliable SEDs for O-stars require a non-LTE treatment.
Figures~9a--c dramatically illustrate how sensitive 
\ion{H}{2} region model predictions of these
ionic abundance ratios are to the
ionizing SED input to nebular plasma simulations. 

	We list for each model the (\teff\ in kK, log~$g$)-pair to identify it.
There are other parameters for a stellar atmosphere model
such as the elemental abundance mix and those that describe
stellar winds, which are treated in each of these non-LTE codes.
Basically, all the atmospheres use solar abundances.
For the \ion{H}{2} region models calculated with 
Pauldrach~et~al.\ (2001) atmospheres, 
the solid line connects models with ``dwarf" atmospheres and the 
dashed line connects models with ``supergiant" atmospheres.
Proceeding from the hot to the cool end,
the Pauldrach~et~al.\ dwarf set has 
(50, 4), (45, 3.9), (40, 3.75), (35, 3.8), and (30, 3.85)
while the supergiant set has
(50, 3.9), (45, 3.8), (40, 3.6), (35, 3.3), and (30, 3).
In several instances, the loci are cut off at the
edges of the plot at the cool end as they track toward
the point computed with the coolest atmosphere.
The Sternberg et~al.\ (2003) paper also uses Pauldrach's WM-BASIC code.
At a given  \teff\,
we have used their model with the smallest log~$g$ in order to
be closest to the supergiant case.
Because the locus using these Sternberg et~al.\ atmosphere models
is for the most part similar to the Pauldrach~et~al.\ supergiant locus, 
we do not show it in Figures~9  to avoid clutter.

	The violet lines with inverted triangles are for \ion{H}{2} region 
models calculated with Lanz \& Hubeny (2003) atmospheres (TLUSTY code).
The solid line connects models using atmospheres with
(45, 4), (40, 4), (35, 4),  and (30, 4) while
the dotted line connects models using atmospheres with
(45,  3.75), (40, 3.5), (37.5, 3.5), (35, 3.25), (32.5, 3.25), and (30, 3).
The orange squares are the results of our
nebular models with the atmospheres in Martins~et~al.\ (2005) 
that use Hillier's CMFGEN code.
The dotted line connects models using atmospheres with
(48.53, 4.01), (42.56, 3.71), (40, 3.5),  (37.5, 3.5), (35,  3.25),
(35, 3.5), and (32.5,  3.5).
The brown line with triangles result from the models using Kurucz (1992)
atmospheres with
(45, 4.5), (40, 4.5), (37, 4), (35, 4), and (33, 4).

	To compare our data with the models in Figures~9a,b, 
we need to divide the observed 
Ne$^{++}$/S$^{++}$
and
Ne$^{++}$/S$^{3+}$
ratios by an assumed Ne/S abundance ratio.
For this purpose,  we adopt a constant Ne/S~= 14.3, the Orion Nebula 
value  (Simpson~et~al.\ 2004).
The open red circles are our prior results for the M83
\ion{H}{2} regions.
The green stars are the M33 results
(adjusted by the assumed Ne/S)
derived from our observed line fluxes
using $N_e$ of 100~cm$^{-3}$.
While the $<$Ne$^{++}$$>$/$<$Ne$^+$$>$ ratio in Figure~9c
has the advantage of
being independent of elemental abundance ratios, it appears to be 
more sensitive to the {\it nebular} parameters than the others,
as will be discussed later.
The trends of the ionic ratios established from the prior M83 study
are remarkably similar, but continued to higher ionization with the 
present M33 objects.
There are two sources, one in M33 and one in M83,
that are deviantly low compared with
the theoretical tracks and the other data.
The point that is low in all three panels  
for M33 is BCLMP~702,
which is the faintest of the M33 regions in the 10.5 and 12.8~$\mu$m 
lines.
Also the FWHM of 903~km~s$^{-1}$ measured
for the 15.6~$\mu$m line is the largest (see Table~2)
and possibly suspect.
The deviant object in M83
(in all three panels) is source RK~268,
which is the faintest of the M83 regions in the 12.8, 15.6, and 18.7~$\mu$m
lines.
For the 10.5~$\mu$m line, it is the second broadest line,
FWHM of 809~km~s$^{-1}$, of any line we measured
in M83 (see Table 2 in R07).
Thus we consider the position of both of these sources
to be duly suspect in Figures~9.
On the whole, the data for both galaxies in panels  a and b 
lie closest to the theoretical loci obtained with
the Pauldrach~et~al.\ supergiant SEDs.
This is particularly notable in Figure~9b, where the other model 
loci are nearly perpendicular to the data point trend in the 
vicinity of where they intersect the data points.
On the other hand, the data in panel c, for the most part, appear to 
lie closer to Martins~et~al., Lanz \& Hubeny~et~al., and
Pauldrach~et~al.\ {\it dwarf} loci.

	The nebular models used to generate Figures~9
are all constant density, ionization-bounded, spherical models.
We used a constant total nucleon density (DENS)  of 1000~cm$^{-3}$
that begins at the star.
Each model used a total number of Lyman continuum photons~s$^{-1}$
($N_{Lyc}$)~= 10$^{49}$.
The same {\it nebular} elemental abundance set was used for all 
nebular models.
We use the same ``reference" set as in Simpson et~al.\ (2004) because in
that paper  we were studying the effects of various  SEDs  on
other ionic ratios and other data sets.
Ten elements are included with their abundance  by number
relative to H as follows:
(He, C, N, O, Ne, Si, S, Ar, Fe) with
(0.100, 3.75E$-$4, 1.02E$-$4, 6.00E$-$4, 1.50E$-$4, 2.25E$-$5, 1.05E$-$5, 
3.75E$-$6, 4.05E$-$6), respectively.
We continue with the same set of abundances as in the M83 paper (R07).
As we mentioned there, 
the set of abundances used is roughly a factor of 
1.5 higher than for Orion and not drastically different from solar.
These heavy element abundances are too high for M33 and
we investigate below how the theoretical  loci will change
when these abundances are scaled  back accordingly.

	We have investigated the effects of changing DENS, $N_{Lyc}$, 
and allowing for a central  evacuated cavity, 
characterized by an  initial radius (R$_{init}$)
before the stellar radiation encounters nebular material.
We term these shell models.
In Figures~10a--c, the resulting changes to Figures~9a--c
are  shown for 12 nebular models run using the 
Pauldrach~et~al.\ (2001) supergiant atmospheres
with \teff\ 35, 40, and 45~kK.
These models are listed along with the symbol in Table~4.
The original Pauldrach~et~al.\ (2001) supergiant locus
(the dashed line connecting the filled circles)
and points derived from the {\it Spitzer} M33 and M83 data are shown again.

	The points  for models 3, 7, and 11 are nearly identical to the 
original points for the Pauldrach supergiant models at the same 
respective \teff.
This can be understood in terms of the ionization
parameter ($U$), which is very useful for gauging ionization
structure.
An increase in $U$ corresponds to  higher ionization (for a given
$T_{\rm eff}$). 
For an ionization-bounded, constant density case,
$$U = [N_e N_{Lyc} (\alpha - \alpha _1)^2/(36 \pi c^3)]^{1/3}~~,$$
\noindent
where ($\alpha$ - $\alpha$$_1$) is the recombination rate coefficient to 
excited levels of hydrogen, and $c$ is the velocity of light
(see Rubin~et~al.\ 1994,  eq.\ 1 and adjoining discussion).
Because\break
($\alpha$ - $\alpha$$_1$)~$\simeq$ 
4.10$\times$10$^{-10}~T_e^{-0.8}$ 
cm$^{3}$~s$^{-1}$ (Rubin 1968 fit to Seaton 1959),
there is only a weak dependence of $U$ on $T_e$ ($\sim$ $T_e^{-0.5}$).
When $U$ is similar, as is the case here with the product of 
$N_e\times$$N_{Lyc}$, the ionization structure is similar.
	With regard to the three shell models  in Figures~10,
the  Str\"omgren radius is $\sim$0.74~pc.
Thus the radial thickness of the shell is slightly less  than half
the radius of the central cavity.  From the visual appearance of our
target \ion{H}{2} regions,
it is unlikely that the theoretical loci need to be tracked to higher
dilutions.

	Another {\it nebular} parameter that can alter the 
theoretical tracks is the set of elemental abundances used.
To investigate this effect and to have a set more representative
of the lower metallicity of M33, we have calculated 
three models (numbers 4, 8,  and 12 in Table~4)
reducing all the heavy element abundances by a factor  of three
from the ``reference" set.
As expected, lower metallicity results in a shift to higher ionization.  
In Figures 10a,b,c the point moves to  the upper right.
For \teff\ 35kK, 
the factors are 
$<$Ne$^{++}$$>$/$<$S$^{++}$$>$~= 1.85,
$<$Ne$^{++}$$>$/$<$S$^{3+}$$>$~= 1.44,
$<$S$^{3+}$$>$/$<$S$^{++}$$>$~= 1.29,  and
$<$Ne$^{++}$$>$/$<$Ne$^+$$>$~= 1.91.
For \teff\ 40kK, 
the respective factors are 
1.39,
1.23,
1.14,  and
1.65.
For \teff\ 45kK, 
the respective factors are 
1.14,
1.10,
1.03,  and
1.45.
Likewise, higher metallicity shifts the point to the lower left (R07).
The above factors show that there is a progression to 
a smaller influence on these ionic ratios with
metallicity as \teff\ increases from 35 to  45kK. 
Additionally, the largest change is always in the
$<$Ne$^{++}$$>$/$<$Ne$^+$$>$  ratio and the
smallest is in the 
$<$S$^{3+}$$>$/$<$S$^{++}$$>$ ratio.
While the $<$Ne$^{++}$$>$/$<$Ne$^+$$>$ ratio has the advantage of
being independent of elemental abundance ratios, it appears to be 
more sensitive to the {\it nebular} parameters than does the 
other ratios.
Note that the y-axis scales are different
in Figures 10 a, b, and c and thus a simple visual 
guide from  the length of the various line segments cannot be
used as an accurate measure of the effect of  changing
a nebular parameter between the three panels.
It is interesting that there appears to be a small, upward shift 
of the M33 star points with respect to the trend of the M83
points (circles), particularly noticeable in Fig.~10c
(and in the electronic colour version).
This is qualitatively what is expected for the lower metallicity
\ion{H}{2} regions of M33.
The dashed line theoretical locus would move up,
connecting the square points.
Nevertheless, for both the case of the M33 and M83 nebulae, we may conclude
that the predicted spread in Figures 10 due to a  reasonable
uncertainty in nebular metallicity is  far less than that due
to the SEDs of the various stellar atmosphere models.

	The magnitude/direction of changes in Figures 10 due to  
varying the {\it nebular} parameters per Table~4 should be 
roughly similar for the other models shown in Figures 9.
It is also noted that we have not considered matter-bounded nebular models.
These would be higher ionization than the corresponding ionization-bounded 
model.
There is also the effect of a change in the abundances used to 
compute the stellar atmosphere models.
This will change the emergent stellar SED (e.g., Mokiem et~al.\ 2004).
As is the case  for a change in  the nebular model abundance set,
such a modification in the stellar model will alter the shape 
of the SED in the same sense;
that is, a higher metallicity will cause more opacity and soften the
SED, and a lower metallicity will do the opposite.
Mokiem et~al.\ (2004) examined this using CMFGEN stellar models
matching both the nebular and stellar metallicities. 
Their Figure~11 tracks the predicted variation in the 
[\ion{Ne}{3}] 15.6/[\ion{Ne}{2}] 12.8 flux ratio
over a range of 0.1~-- 2~Z$_{\sun}$.
Although beyond the scope of this paper, it would be interesting to compare 
models using different stellar atmosphere metallicities, especially if the 
environment indicates significant departures from solar. 
However, the proper abundances are not
accurately known and comparisons like those in this paper 
will help decipher the proper values.
With the present paper, however, we have been investigating 
which of the models best fits the observations and 
such a comparison works most effectively with a fixed set 
of abundances {\it common to all models}, which are the solar ones.   

	There appears to be a remarkable ``convergence"
of the nebular models close to the hot (right) side of Figures 9.
Depending on the set of stellar  atmosphere  models,
there is a significant  spread in \teff\ ($\sim$42--50kK)
(as  well  as the log~$g$  parameter) in the vicinity where this occurs.
For instance, the results in all  three panels are essentially identical 
using either the TLUSTY (45, 3.75) or the WM-BASIC (50, 3.9) SED.
A likely explanation for this convergence
behaviour in Figures 9 is that at these high \teff\ values,
the ionization balance in the atmospheres shifts to higher ionization stages 
which have fewer lines that can influence the model calculations. 
Thus, blocking and blanketing effects no longer dominate the models as 
strongly, and the line radiation pressure is also less significant.
This means that the influence of the expanding atmosphere decreases. 
As a result of all this, at high \teff\,
the model calculation are no longer as critically dependent on the correct 
treatment of the complex details of the physics involved. 

\section{Discussion}

	With the recent papers by W08
and Lebouteiller et~al.\ (2008), hereafter L08,
we are in a position 
to further examine the Ne/S  ratio discussed in \S4.
Fundamental observational data are vital to test and constrain theories 
of nucleosynthesis and 
GCE. 
A very valuable adjunct would be to find how much the Ne/S
ratio can vary or whether or not there is a fairly ``universal" value.
W08 obtained an average Ne/S~= 12.5$\pm$3.1
from  {\it Spitzer} observations of  13 blue compact dwarf galaxies.
They found no correlation between their  Ne/S ratios with
metallicity (the Ne/H ratios).
With their Table~3 of line fluxes, we apply the same methods 
used here for the M33 objects to derive the various ionic abundance ratios.
For the same blue compact dwarf galaxy, we obtain somewhat higher Ne/S ratios.
We are particularly interested in combining their observations with 
ours to address the Ne/S ratio and whether there is any correlation
with degree of ionization as measured by the Ne$^{++}$/Ne$^+$ and
S$^{3+}$/S$^{++}$ ratios.
Nine of the galaxies in W08
had all four of the necessary
lines detected to  derive both of  these ratios (i.e., we excluded
their 4 objects which had at least one upper limit on a line flux).
In Table~5, we list in columns  2--4,  Ne$^{++}$/Ne$^+$, S$^{3+}$/S$^{++}$,
and Ne/S derived by us using the individual  $T_e$-- and $N_e$--values 
in Wu et~al.'s  Table 2.
In columns 5--7, we list the same  set derived here assuming
$T_e$~= 10000~K  and $N_e$~= 100~cm$^{-3}$.
It is apparent  that there is very little difference between
these due to what ($T_e$, $N_e$) values are used.
The median (average) Ne/S for the 9 galaxies in column 4 is 14.0 (14.9),
while it is 13.9 (14.9) from column 7.
However, for these 9 objects, the median Ne/S is 12.4 and the average 12.7
(from  Table 4 in W08).

L08 reported their Spitzer observations
of three very massive \ion{H}{2} region:
the Galactic source NGC~3603;
the extremely massive 30~Dor in the Large Magellanic Cloud; and
N~66 (NGC~346), the largest in the Small Magellanic Cloud.
With their Table~2 of line fluxes, we apply the same methods 
used here for the M33 objects to derive the various ionic abundance ratios.
In their Table~6, they list derived abundance ratios
for several position observed in each object: 3 for NGC~3603, 
15 for 30~Dor, and 11 for N~66.
Again, we find somewhat higher Ne/S ratios at all positions than 
the corresponding value obtained from their Table~6.
In Table~6 here, we list in columns  2--4,  Ne$^{++}$/Ne$^+$, S$^{3+}$/S$^{++}$,
and Ne/S derived by us using the same $T_e$-- and $N_e$--values 
as they had used.
These are 10000~K  and 1000~cm$^{-3}$ for NGC~3603; 
10000~K  and 100~cm$^{-3}$ for 30~Dor;  and
12500~K  and 100~cm$^{-3}$ for N~66.
The median Ne/S ratios we derive for NGC~3603, 30~Dor, and N~66.
are 14.6, 11.4, and 10.1, respectively.

	Some variation is traceable to the adoption of different 
[\ion{S}{3}] effective collision strengths.
We  use the set by Tayal \& Gupta (1999);
W08 and  L08
stated that they used values from the IRON Project,
which  for this ion would be  ``paper X" (Galav\'\i s et~al.\ 1995).
There is a  substantial difference between the values in these papers.
For instance the values between  the ground-state, fine structure levels
(where  the  [\ion{S}{3}] IR lines arise)
at $T_e$~=  10000~K are: 3.98, 1.31, and 7.87 (Tayal \& Gupta)
and 2.331, 1.110,and 5.411 (Galav\'\i s et~al.) for transitions
$^3P_0$--$^3P_1$, 
$^3P_0$--$^3P_2$, and
$^3P_1$--$^3P_2$  respectively.
Furthermore, these three values show a change in
{\it opposite directions} with $T_e$ decreasing from 10000~K.
The above three collision strengths increase in the  
Tayal \& Gupta work but decrease in the other study.
In  \ion{H}{2} regions, S$^{++}$ is almost always more abundant than
S$^{3+}$ as is the case for every point here except 
1 of the 9 
W08 galaxies addressed here.
Because we  use the larger [\ion{S}{3}] effective collision strengths,
the S$^{++}$ abundance we infer will be lower than what 
was derived in  the other two papers.
This will then  result in the total S abundance (S$^{++}$ + S$^{3+}$) 
being smaller and hence the Ne/S abundance larger than what  
W08 and L08
found.
Following Simpson et~al.\  (2007), we repeat here the 
references for the other effective collision strengths we use:
[\ion{S}{4}] (Saraph \& Storey 1999),
[\ion{Ne}{2}] (Saraph \& Tully 1994),
and 
[\ion{Ne}{3}] (McLaughlin \& Bell 2000).
Apparently 
W08 and L08
did use the same sets for [\ion{Ne}{2}] 
and [\ion{S}{4}] but probably used 
``paper V" (Butler \& Zeippen 1994) from the IRON Project
for [\ion{Ne}{3}].
We note that near $T_e$~= 10000~K, the  values for  
[\ion{Ne}{3}] are quite similar.
Even without passing judgment on the relative merits of one
set of effective collision strengths vs.\ another,
it is crucial for the purposes of  this paper that we  continue
the analysis  with a homogeneous treatment.
For this we  will use the adjusted ionic ratios from columns 2--4
in Tables  5 and 6.

	In Figure~11, we plot Ne/S vs.\ Ne$^{++}$/Ne$^+$ 
for our M33 results 
(star symbol) using the same 22 sources mentioned in  Fig.~6.
The linear least-squares fit to the M33 points  is,

~~~~~~~~~~~~~~~Ne/S~= 17.02$\pm$0.63~$-$~(0.78$\pm$0.50)~Ne$^{++}$/Ne$^+$,

\noindent
which is shown as the solid line.
The negative gradient is not statistically significant.
The results from our prior M83 study (R07)
are shown as circles.
The wide spread in the M83 points dramatically
demonstrates the limitations of our method to infer Ne/S
[except as an upper limit] 
when Ne$^{++}$/Ne$^+$ is low (i.e., for low ionization objects).
The squares show the 
W08 data for the 9 blue compact dwarf galaxies, 
as discussed above.
The linear least-squares fit to those 9 points  is,

~~~~~~~~~~~~~~~Ne/S~= 18.17$\pm$1.07~$-$~(0.93$\pm$0.24)~Ne$^{++}$/Ne$^+$.

\noindent
While our M33 data indicate no  statistically  significant slope,
the fit to the 9 blue compact dwarf  galaxies shows a
statistically  significant decrease with higher Ne$^{++}$/Ne$^+$
(the dotted line in Fig.~11).
We note that the median (average) Ne/S for the 9 galaxies is 14.0 (14.9), 
very close to  the Orion value of 14.3  shown as the dashed line.
If we exclude the four lowest ionization galaxies and fit just
the five with the highest ionization, then

~~~~~~~~~~~~~~~Ne/S~= 15.09$\pm$1.15~$-$~(0.49$\pm$0.20)~Ne$^{++}$/Ne$^+$,

\noindent
and this flatter negative slope (2.5~$\sigma$) is  no longer 
statistically significant.

	In Fig.~11, we also show the reanalyzed results of L08 (see Table~6). 
There appears to be a trend with the three positions in NGC~3603 
(red  asterisks) having the highest median Ne/S, followed by
the 15 positions in 30~Dor (green triangles),
and the 9 positions in N~66 (blue diamonds) having the lowest median
ratio.
This trend is correlated with the metallicity  with
NGC~3603 approximately solar,  30~Dor $\sim$0.7 solar, and
N~66 $\sim$0.2 solar (L08 and references therein).  
No error bars are shown on these points, but they may be expected  to
be roughly 20~percent according to L08.  A picture that is consistent
with the data in Figure~11 is that  at  the lowest ionizations 
there remains considerable S still present at these positions in
the form of S$^+$, molecular gas, and dust.
The three 30~Dor positions with the lowest ionization,
as measured by both Ne$^{++}$/Ne$^+$  and S$^{3+}$/S$^{++}$,
are \#10 and 17, followed by \#8 in Table~6.  
An inspection of Fig.~2 in L08 shows that positions
\#10 and 17 are near the periphery of their 30~Dor image
and thus  might be expected  to be characterized by  lower  ionization 
than interior positions.
The linear least-squares fit to the data in Table~6 excludes these
three lowest ionization 30~Dor points as well as the NGC~3603 points.
The dash-dot line fits 23 positions, those in N~66 and the 12 
remaining in 30~Dor, with the highest ionization.
We find,

~~~~~~~~~~~~~~~Ne/S~= 12.08$\pm$0.41~$-$~(0.37$\pm$0.12)~Ne$^{++}$/Ne$^+$,

\noindent
with a slope that is marginally significant (3.1~$\sigma$).
We note that there is no basis to extrapolate the line
to the edges of the graph as shown, and indeed no theoretical basis 
for a straight line.  What we would expect once all other S species
become negligible with respect to S$^{++}$ and S$^{3+}$ 
is that an asymptotic Ne/S will be approached at the higher ionizations.

	Similar fits to Ne/S vs.\ S$^{3+}$/S$^{++}$ are shown in Fig~12
and result in,

~~~~~~~~~~~~~~~Ne/S~= 16.67$\pm$0.71~$-$~(3.11$\pm$5.41)~S$^{3+}$/S$^{++}$, 

\noindent
for  22 sources in M33.
The slope is not significant statistically.
For the 9 blue compact dwarf galaxies,

~~~~~~~~~~~~~~~Ne/S~= 17.70$\pm$1.14~$-$~(6.39$\pm$2.00)~S$^{3+}$/S$^{++}$. 

\noindent
This indicates a marginal, statistically  significant 
(3.2~$\sigma$) negative slope as S$^{3+}$/S$^{++}$ increases.
As with Fig.~11, if we exclude the four lowest ionization galaxies 
and fit just the five with the highest ionization, then,

~~~~~~~~~~~~~~~Ne/S~= 14.42$\pm$1.30~$-$~(2.84$\pm$1.72)~S$^{3+}$/S$^{++}$. 

\noindent
This flatter negative slope (1.6~$\sigma$) is  no longer 
statistically significant.
The linear least-squares fit to the reanalyzed L08 data in Table~6 
for the same 23 positions as done in Fig.~11 results  in,

~~~~~~~~~~~~~~~Ne/S~= 12.14$\pm$0.44~$-$~(5.69$\pm$1.88)~S$^{3+}$/S$^{++}$, 

\noindent
with a slope  that is borderline significant (3.0~$\sigma$).

	It is tantalizing to conjecture that 
for high ionization objects, all the dominant states of Ne and S are 
measured, resulting in a robust estimate of the true Ne/S abundance ratio.
For objects as diverse as the Orion Nebula and NGC~3603, 
the M33 \ion{H}{2} regions, 30~Dor, N~66, and the blue compact 
dwarf galaxies, there is remarkably little variation in the Ne/S derived.
We note that if we were to adopt the smaller
[\ion{S}{3}] effective collision strengths
(Galav\'\i s et~al.\ 1995) that 
W08 and L08 used,
all values for Ne/S that we have derived would be lower,
including that  for  Orion.
A rough estimate of how  much lower is $\sim$2
(that is, $\sim$15--25~percent), which can be surmised from
the comparison with the 
W08 and L08
Ne/S values and the recalculated
values here in Tables 5 and 6.
All of these Ne/S values point to a much larger ratio than 
the ``canonical" solar  value of $\sim$5 (see next section)
and what had previously been predicted by the GCE models 
mentioned in the Introduction.
The question, of how universal the Ne/S ratio may be, 
will require further study.
Two of the blue compact dwarf galaxies (UM461 and IIZw40)
with the highest degree of  ionization strongly affect the
line fits in Figures 11 and 12 and the conclusion that 
Ne/S continues to  decrease, reaching a value of 10.
It is possible that it does take such a high degree of ionization
to percolate off any substantial amount of S that may still be
tied up in grains, molecules, and  S$^+$.
Further {\it  Spitzer} observations of high-ionization
\ion{H}{2} regions will be useful for this purpose.
We note that the type of analysis 
done here will not hold for even higher ionization objects where
substantial amounts of Ne exist as Ne$^{3+}$  or higher and/or
S as S$^{4+}$ or higher.

\section{Summary and conclusions}

	We have observed emission lines of 
H(7--6) 12.37,
[\ion{Ne}{2}] 12.81,
[\ion{Ne}{3}] 15.56,
[\ion{S}{3}] 18.71, and [\ion{S}{4}] 10.51~$\mu$m
cospatially with the {\it  Spitzer Space Telescope} using the Infrared
Spectrograph (IRS) in short-high mode (SH).  From the measured fluxes,
we estimate the ionic abundance ratios Ne$^{++}$/Ne$^+$,
S$^{3+}$/S$^{++}$, and S$^{++}$/Ne$^+$ 
in 25 \ion{H}{2} regions in the substantially face-on spiral galaxy M33.  
These nebulae cover a range from 0.71 to 6.73~kpc
in deprojected galactocentric distance  R$_G$.
We find a correlation of increasingly higher ionization with increasing R$_G$.
This is seen in the variation of 
Ne$^{++}$/Ne$^+$,
S$^{3+}$/S$^{++}$,
and $<$S$^{++}$$>$/$<$Ne$^+$$>$ with R$_G$ (see  Figures 3--5).  
A possible reason may be due to the lower metallicity at larger R$_G$
causing the exciting stars to have a harder ionizing spectrum.  
As mentioned in the introduction and discussed in considerable
detail (e.g., Morisset~et~al.\ 2004), there are other
effects that could mimic this.
We find a 
decrease in the metallicity, as measured by the Ne/H and S/H ratio,
with increasing R$_G$ (see  Figures 7--8).  
The linear-log gradients  that we derive are in 
remarkable agreement  with  the recent
value  for the log~(O/H) gradient of  
$-$0.054$\pm$0.011~dex~kpc$^{-1}$ (Magrini~et~al.\ 2007)
derived from  optical observations  of 14 \ion{H}{2} 
regions in  M33.
Because their slope is practically identical to ours 
for  log~(Ne/H), we infer a Ne/O ratio of  0.28.

	By sampling the dominant ionization states of Ne and S
for \ion{H}{2} regions, we can approximate the Ne/S ratio
by (Ne$^+$ + Ne$^{++}$)/(S$^{++}$ + S$^{3+}$).
For M33, we find no significant variation in the Ne/S ratio with R$_G$.
Both Ne and S are the products of $\alpha$-chain reactions
following carbon and oxygen burning in stars,
with large production factors from core-collapse supernovae.
Both are primary elements, making their yields depend very
little on the stellar metallicity. 
Thus, at least to ``first order", it is expected that Ne/S remains 
relatively constant throughout a galaxy.
As discussed in \S4 and \S7, our estimate for Ne/S
has accounted for neither the presence of S$^+$
nor S that may be tied up in grains or molecular gas.

The data presented here for M33, combined with other {\it Spitzer} data
(see Figures 11 and 12), are consistent with our view that there are
now reliable estimates for the {\it total} Ne/S ratio.
As long as the degree of ionization is sufficiently high such that
the amount of sulfur in forms other than S$^{++}$ and S$^{3+}$ is small,
the methodology used here will provide a robust total Ne/S estimate. 
The median Ne/S value we derive for the M33 objects is 16.3,
but this includes many with relatively low ionization.
Although there is no statistically significant gradient with Ne$^{++}$/Ne$^+$,
two of the sources, BCLMP~638 and 280, with the highest ionization 
are among those with the lowest Ne/S at  13.9 and 13.3, respectively.
For the W08 blue compact dwarf galaxies, we find a median Ne/S for 
the 9 galaxies of 14.0.
When we consider just the 5 with the highest ionization, the median drops to 13.2.
 From Table~6 with the recomputed Ne/S from the L08 observations,
we find a median Ne/S of 10.1 for N~66 and 11.4 for 30~Dor.
This median for 30~Dor drops to 11.3 if the three lowest ionization values
are not included.
 Although the data are limited, we note the possibility that the true 
Ne/S ratio may be less for lower metallicity galaxies.
N~66 has the lowest metallicity ($\sim$0.2 solar) while 
30~Dor in the LMC 
has an intermediate  value  compared with the  Milky  Way.
On the other hand, W08 found no correlation of Ne/S with 
metallicity for the blue compact dwarf galaxies.

	The solar abundance, particularly of Ne, remains the subject of 
much controversy (e.g., Drake \& Testa 2005; Bahcall, Serenelli, \& Basu 2006;
and references in each of these).
While we cannot directly address the solar abundance with our
observations of extragalactic \ion{H}{2} regions,
it is important to  have reliable benchmarks for the Ne abundance.
There appears to be a growing  body of evidence that the Ne abundance
[its fractional number abundance relative to H
log~H~= 12, by definition and termed $A$(H)]
is substantially higher in the solar neighborhood, and even in the Sun
itself, than the ``canonical" solar values given in two often-referenced papers.
These  papers have for the Sun:  $A$(Ne)~= 7.87, $A$(S)~= 7.19  (Lodders 2003)
and $A$(Ne)~= 7.84, $A$(S)~= 7.14  (Asplund, Grevesse,  \& Sauval 2005).
Thus according to both, Ne/S~$\sim$5. 
It is now generally accepted that Ne has the least
well determined solar abundance among the most abundant elements.
One of the proponents for a higher neon abundance
pointed out that an $A$(Ne)~= 8.29 would reconcile solar models with 
the helioseismological  measurements (Bahcall, Basu, \& Serenelli 2005).
Using this value  together with the $A$(S) values above,
we obtain Ne/S of 12.6 and 14.1,  respectively,
close to the Orion Nebula ratio 14.3 (Simpson~et~al.\ 2004) 
and the various median ratios that range from 10.1 to 16.3 derived earlier.

	As discussed in the last section,
there is a surprisingly large difference between the effective collision
strengths calculated for [\ion{S}{3}].
It  would be very  worthwhile to resolve the differences  with  a new
calculation or possibly experimental  work.
This is  a very important matter because the dominant sulfur ionization 
state in the preponderance of \ion{H}{2} regions is  S$^{++}$.
Another desirable avenue for future work would be 
ground-based optical observations that could
cover the S$^+$ ion, which must be  present at some  level,
and which cannot be done with {\it Spitzer}.
A programme designed to cover [\ion{S}{2}] 6716,31~\AA\
as well as the [\ion{S}{3}] 9069,9531~\AA\ lines,
particularly for the regions we deem high ionization here,
would confirm whether or not the S$^+$ abundance is
negligible and the reliability of our Ne/S values.

	In our earlier M83 paper (R07), we discussed
what the Ne/S ratio was predicted to be 
according to calculations based on the theoretical  nucleosynthesis, galactic
chemical  evolution models  of Timmes, Woosley,  \& Weaver (1995).
The ratio was about 3.8.
Since then there have been improved models (Woosley \& Heger 2007).
Their calculation considers massive stars from 12 to 120~M$_\odot$
starting with the solar abundance set of Lodders (2003).
One difference from the Timmes~et~al.\ work
is the addition now of a treatment for  mass loss.
This  includes mass loss on the main sequence as well  as red giant,
and Wolf-Rayet phases. 
It also includes improvements in the explosion physics and
nuclear cross sections.
Stellar rotation is not included and according to
Stan Woosley (private communication), will have an uncertain effect.
  Based on Fig.~7 in Woosley \& Heger (2007), the Ne/S ratio
is predicted to be $\sim$8.6.
These models are based on starting with the solar abundances.
Since several of  the objects observed by {\it Spitzer} are
lower than solar metallicity, it would be very interesting to
compute the Ne/S nucleosynthesis yields starting with lower metallicities.

	The data set here, combined with our previous M83 results (R07), 
may be used as constraints on the ionizing SEDs for the stars exciting 
these nebulae by comparing the ratio of fractional ionizations
$<$Ne$^{++}$$>$/$<$S$^{++}$$>$, 
$<$Ne$^{++}$$>$/$<$S$^{3+}$$>$,
and
$<$Ne$^{++}$$>$/$<$Ne$^+$$>$
vs.\ 
$<$S$^{3+}$$>$/$<$S$^{++}$$>$
with predictions
made from our photoionization models using stellar atmosphere models
from several different sources.
In Figures~9a,b we show the comparison assuming 
that the Ne/S ratio does not vary and equals the Orion Nebula value.
Generally, the best fit is to the nebular models using the
supergiant stellar atmosphere models 
(Pauldrach~et~al.\ 2001)
computed with the WM-BASIC code.
The comparison shown in Figure~9c is independent of the Ne/S ratio.
For the most part, the Ne$^{++}$/Ne$^+$ values appear to lie closer 
to the theoretical loci that use the Martins~et~al., Lanz \& Hubeny~et~al., 
and Pauldrach~et~al.\ {\it dwarf} SEDs.
While the Ne$^{++}$/Ne$^+$ 
ratio has the advantage of
being independent of elemental abundance ratios, it appears to be 
more sensitive to the {\it nebular} parameters than does the 
other ionic ratios used in Figures 9 and 10.
This fact tends to make it less unique in its ability to discriminate
between the stellar SEDs we present in this paper.
We note that these comparisons are mainly qualitative
since these ionic ratios depend not only on the SED, but also
on the nebular parameters discussed as well
as the effects of the stellar metallicity on the SED.

\begin{acknowledgments}
This work is based on observations made with the 
{\it Spitzer Space Telescope},
which is operated by the Jet Propulsion Laboratory, California Institute
of Technology under NASA contract 1407.  Support for this work was provided
by NASA for this {\it Spitzer} programme identification 20057.
We thank Stan Woosley for providing information 
on the Ne/S ratio from a nucleosynthesis,
galactic chemical evolution perspective.
The referee Christophe Morisset provided comments that helped improve the paper.
Our computer support at NASA Ames was handled very well by David Goorvitch.
We thank Danny Key, Erik  Krasner-Karpen, and  Matt Lattanzi
for assistance with the data reduction.
The nebular models were run on JPL computers, initially a Cray 
and more recently, a Dell Intel Xeon cluster.
Funding for computer use in this investigation was provided by 
the JPL Office of the Chief Information Officer.
We are grateful to Heidi Lorenz-Wirzba for excellent computer support.
\end{acknowledgments}


\begin{figure}
\centering
\resizebox{14.0cm}{!}{\includegraphics{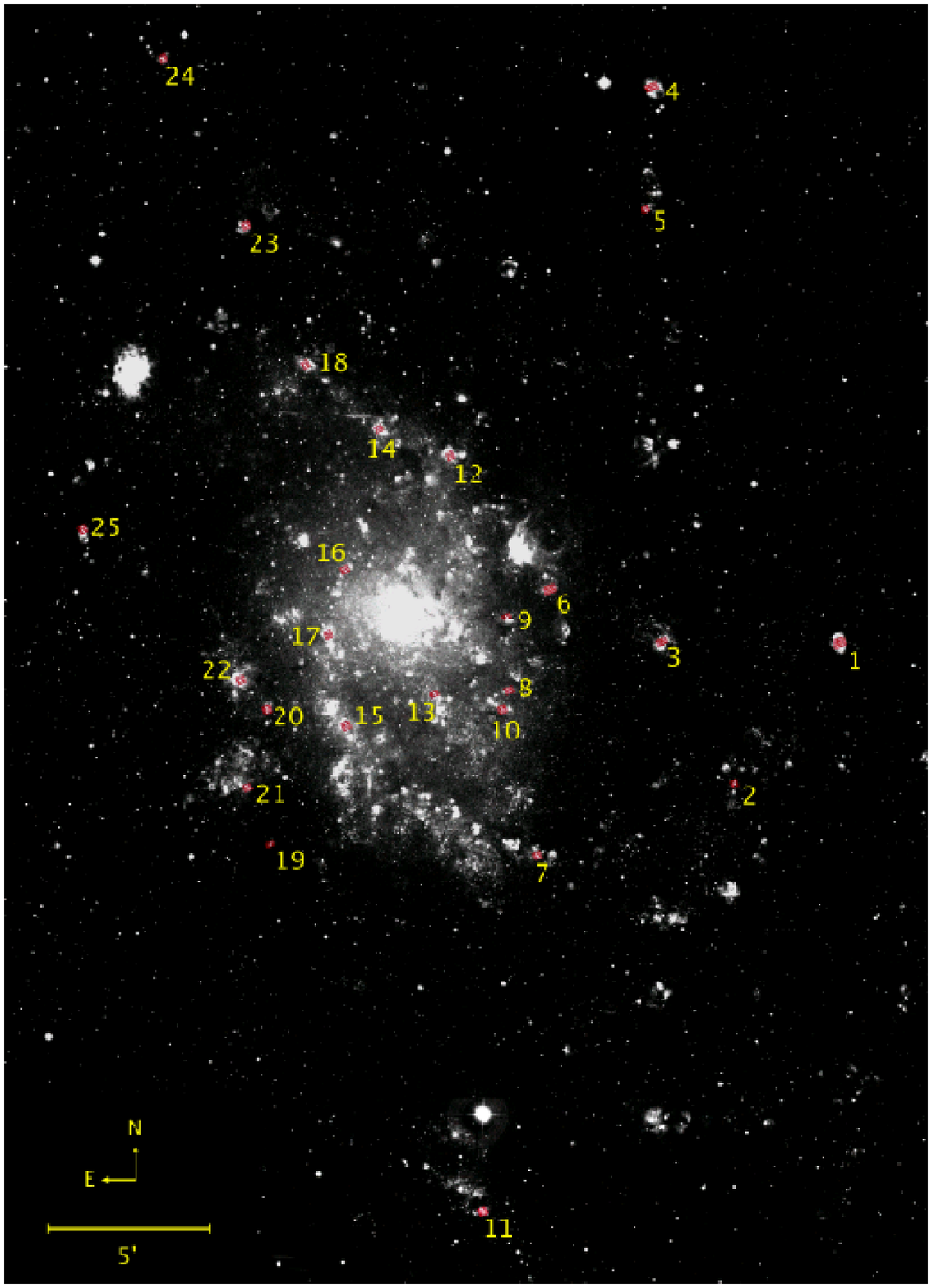}}
\vskip-0.1truein
\caption[]{The positions and apertures observed for 25 \ion{H}{2} regions
are shown superimposed on an H$\alpha$ image of 
the substantially face-on (tilt 56$^{\rm o}$) M33.
The nebulae are numbered W to E (see Table~1).
N is up and E left.}
\end{figure}


\begin{figure}
\begin{center}
\vskip-0.25truein
\begin{tabular}{cc}
\resizebox{70mm}{!}{\includegraphics{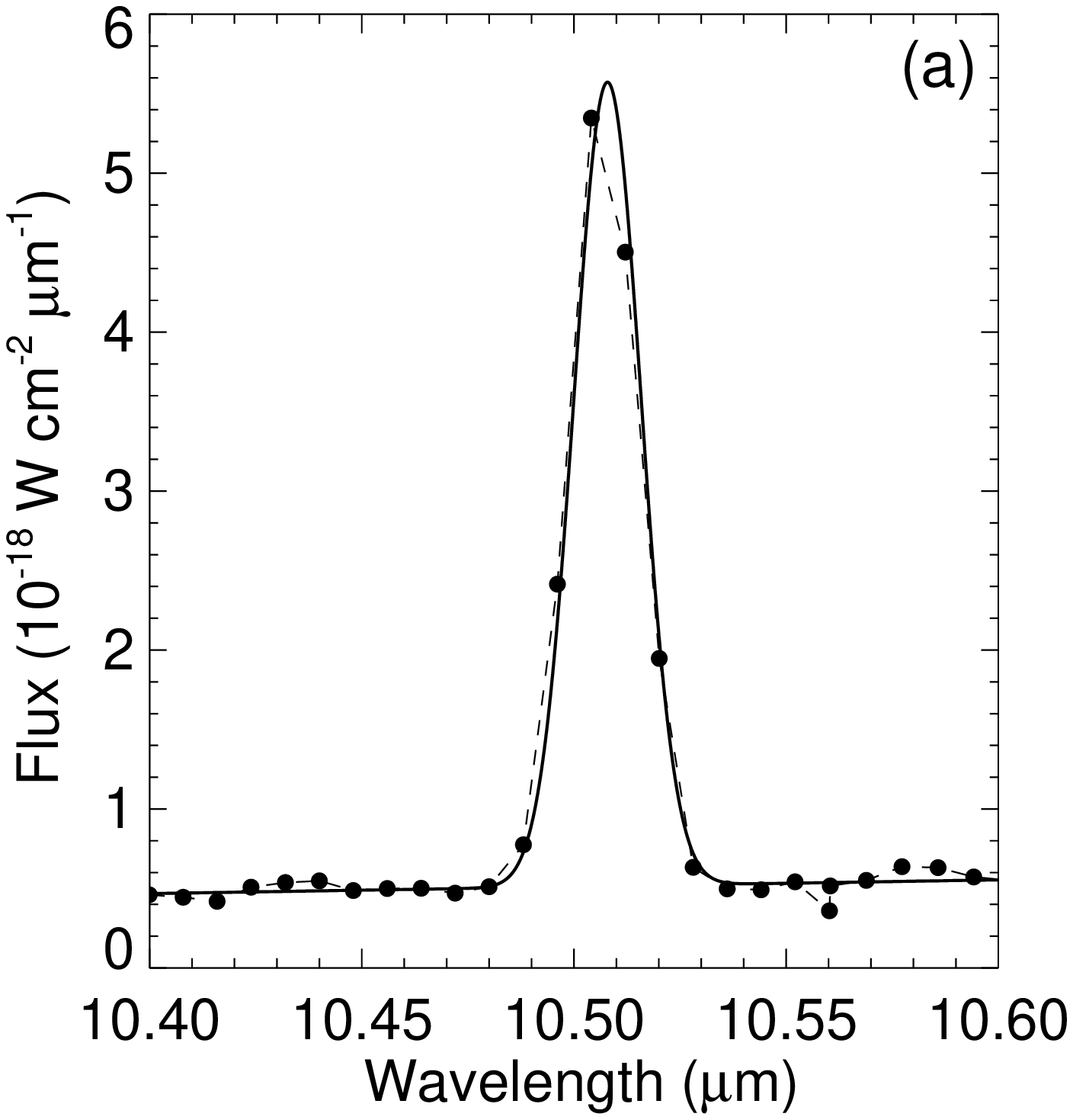}} &
\resizebox{70mm}{!}{\includegraphics{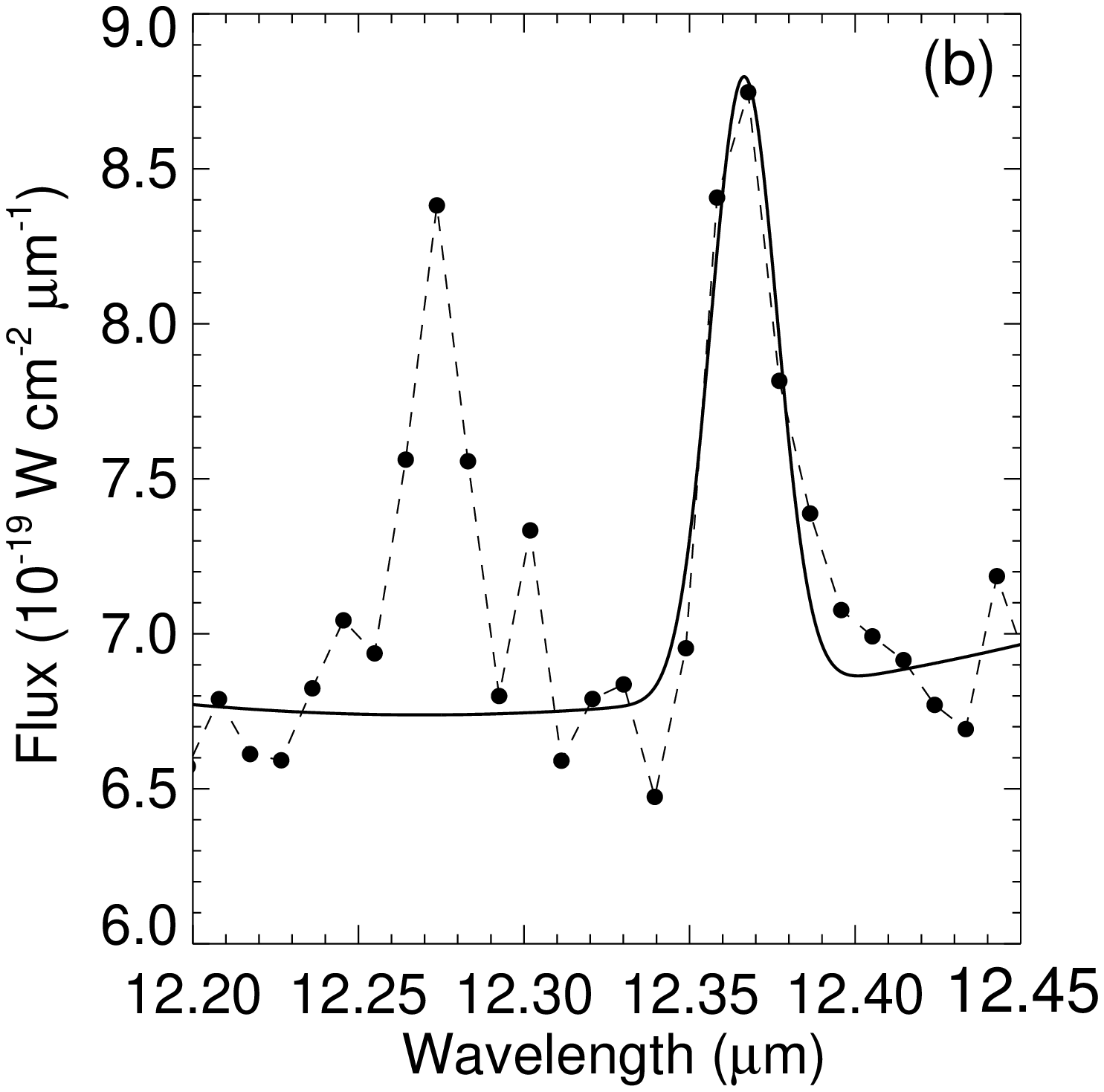}} \\
\resizebox{70mm}{!}{\includegraphics{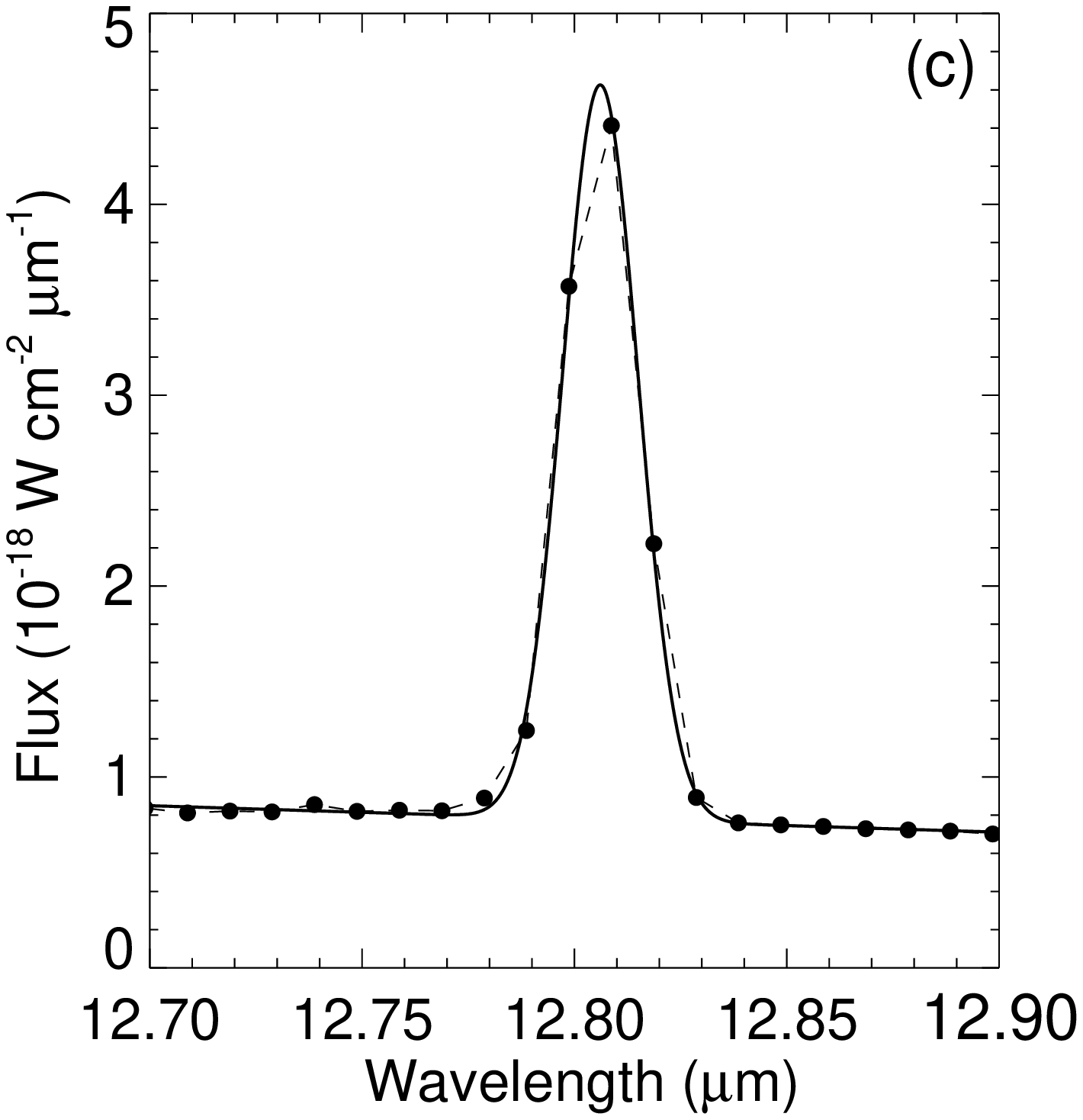}} &
\resizebox{70mm}{!}{\includegraphics{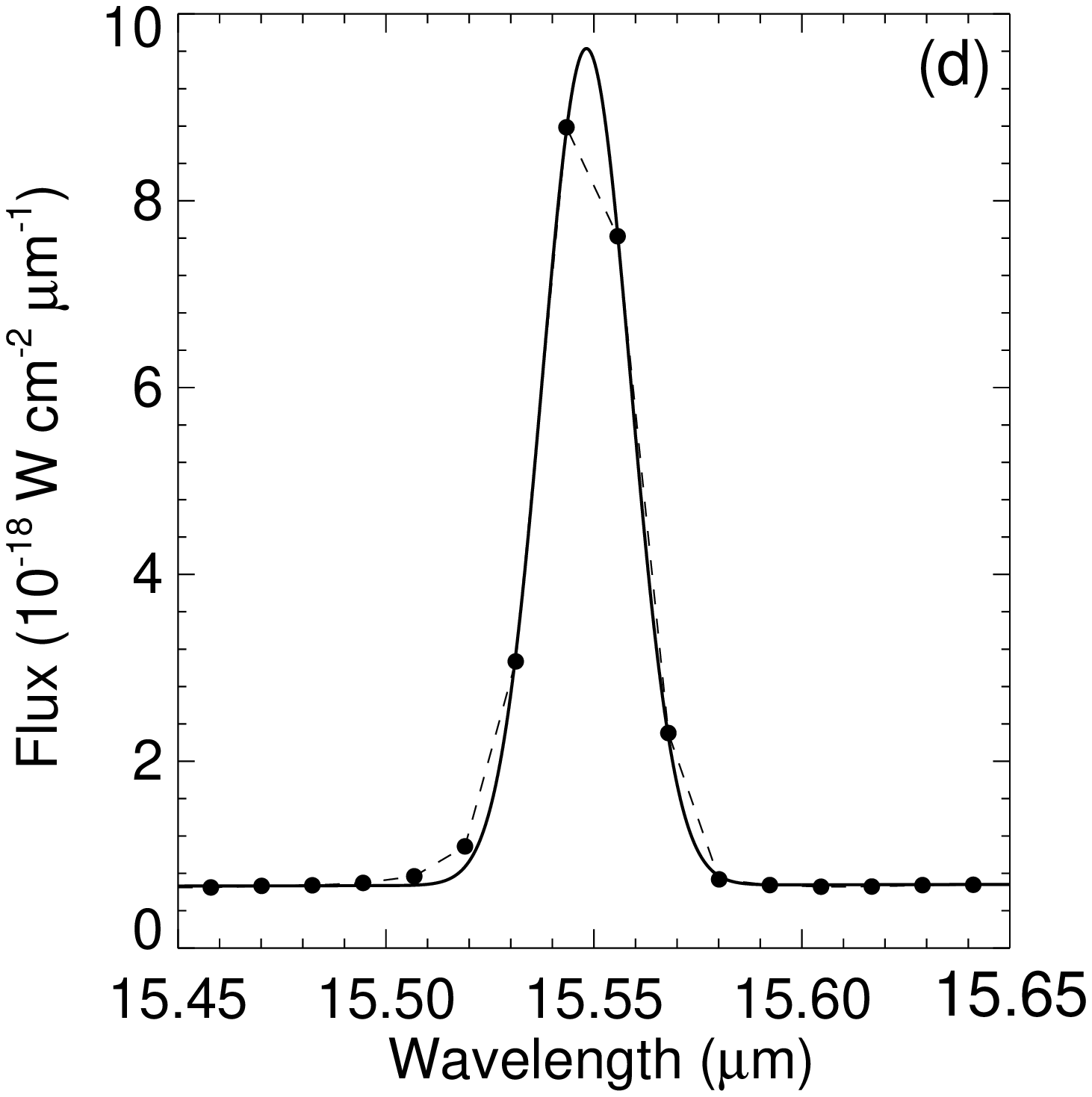}} \\
\end{tabular}

\vskip-0.1truein
\begin{tabular}{c}
\resizebox{70mm}{!}{\includegraphics{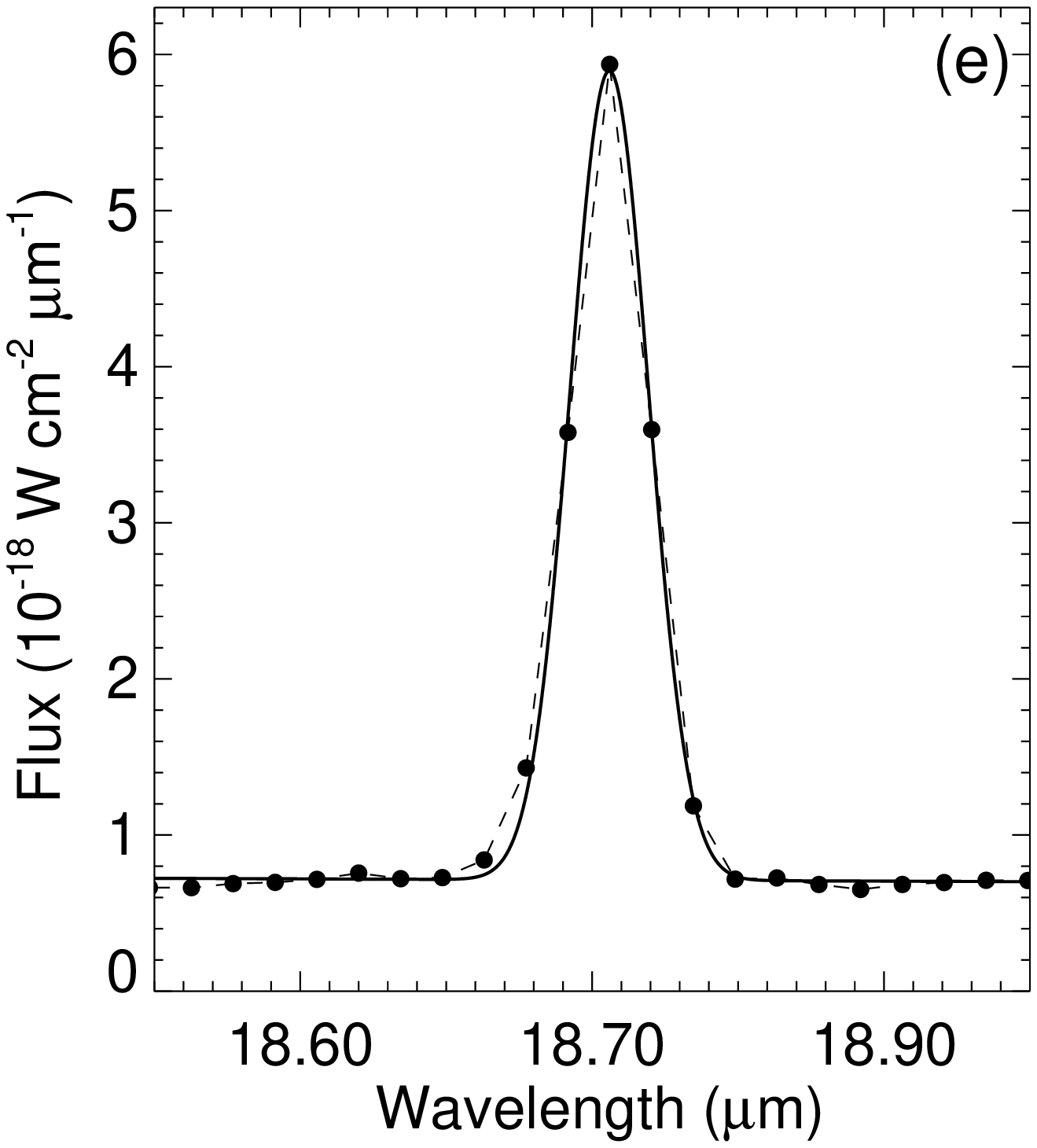}} \\
\end{tabular}
\vskip-0.1truein
\caption{Measurements of the five emission lines in the \ion{H}{2} region
         BCLMP~45 (\#6 in Fig.~1): {\bf (a)} [\ion{S}{4}] 10.5~$\mu$m;  
         {\bf (b)} H(7--6) 12.4~$\mu$m; the feature on the blue side is
         H$_2$~S(2);
         {\bf (c)} [\ion{Ne}{2}] 12.8~$\mu$m;
         {\bf (d)} [\ion{Ne}{3}] 15.6~$\mu$m; and 
         {\bf (e)} [\ion{S}{3}] 18.7~$\mu$m.
         The data points are the filled circles.  The fits to the continuum 
         and Gaussian profiles are the solid lines.  Such measurements provide 
         the set of line fluxes for further analysis.}
  \end{center}
\end{figure}


\begin{figure}
\centering
\resizebox{14.0cm}{!}{\includegraphics{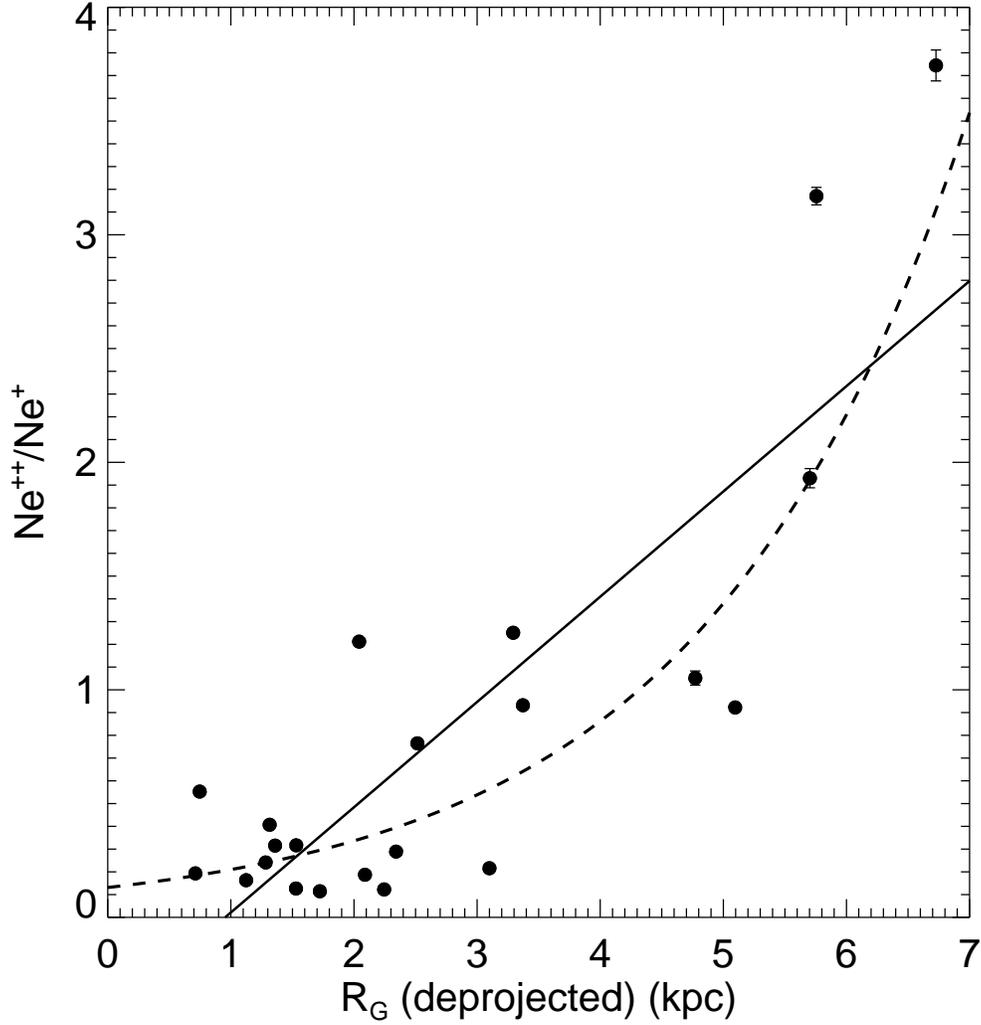} }
\vskip0.1truein
\caption[]{Plot of the ionic abundance ratio 
Ne$^{++}$/Ne$^+$, which is 
derived from the measured line flux ratios for 22 of the
\ion{H}{2} regions,  vs.\ R$_G$.
We assume an electron density ($N_e$) of  100~cm$^{-3}$.
There is extremely little change with $N_e$ over the range
expected for these regions.
The solid line is a linear least-squares fit
while the dashed line results from
a linear least-squares fit to these points in
a log(Ne$^{++}$/Ne$^+$) vs.\ R$_G$ plot.
In both cases, there is a significant positive correlation with R$_G$.
Error bars here and in Figs 4--8 are for the propagated measurement
uncertainties and do not include the systematic uncertainties (see text).}

\end{figure}


\begin{figure}
\centering
\resizebox{14.0cm}{!}{\includegraphics{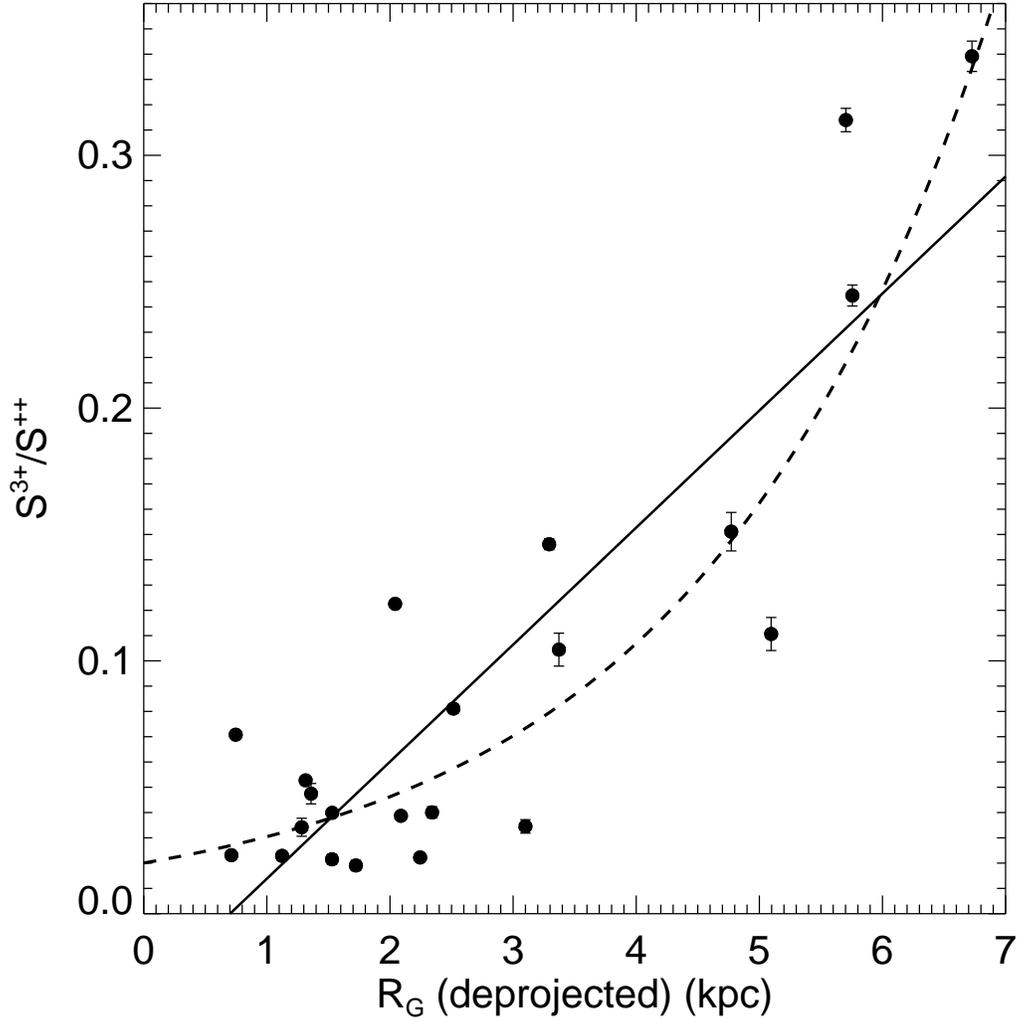} }
\vskip0.1truein
\caption[]{Plot of the ionic abundance ratio 
S$^{3+}$/S$^{++}$ vs.\ R$_G$
for the same 22 \ion{H}{2} regions  as Figure~3.
Here also,  we assume $N_e$ of  100~cm$^{-3}$.
The solid line is a linear least-squares fit
while the dashed line results from
a linear least-squares fit to these points in
a log(S$^{3+}$/S$^{++}$) vs.\ R$_G$ plot.
In both cases, there is a significant positive correlation with R$_G$
(see text).}

\end{figure}


\begin{figure}
\centering
\resizebox{14.0cm}{!}{\includegraphics{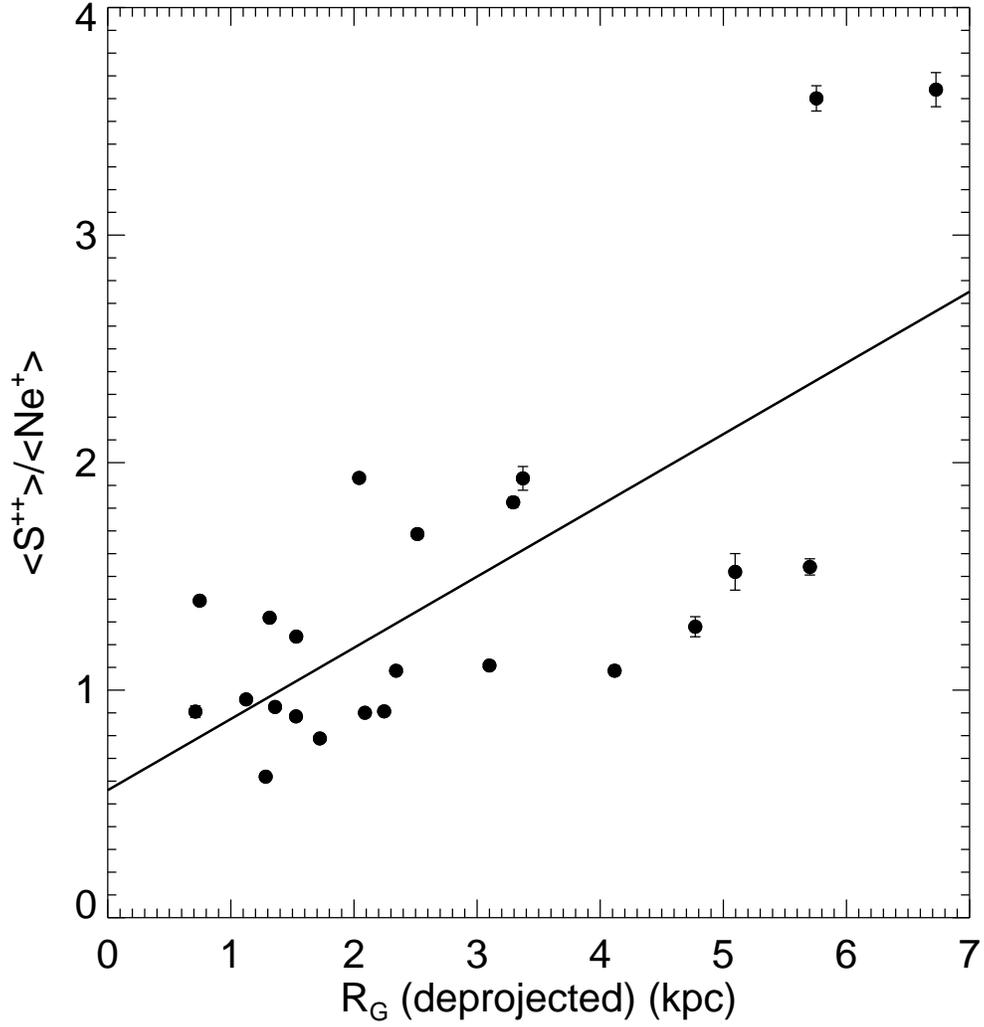} }
\vskip0.1truein
\caption[]{
The fractional ionic abundance ratio  
$<$S$^{++}$$>$/$<$Ne$^+$$>$ vs.\ R$_G$ for 23 \ion{H}{2} regions.
The linear least-squares fit for an assumed $N_e$ of 100~cm$^{-3}$ is shown.
The plotted $<$S$^{++}$$>$/$<$Ne$^+$$>$ ratio assumes an Orion Nebula Ne/S 
abundance ratio of 14.3 (Simpson et~al.\ 2004).}

\end{figure}


\begin{figure}
\centering
\resizebox{14.0cm}{!}{\includegraphics{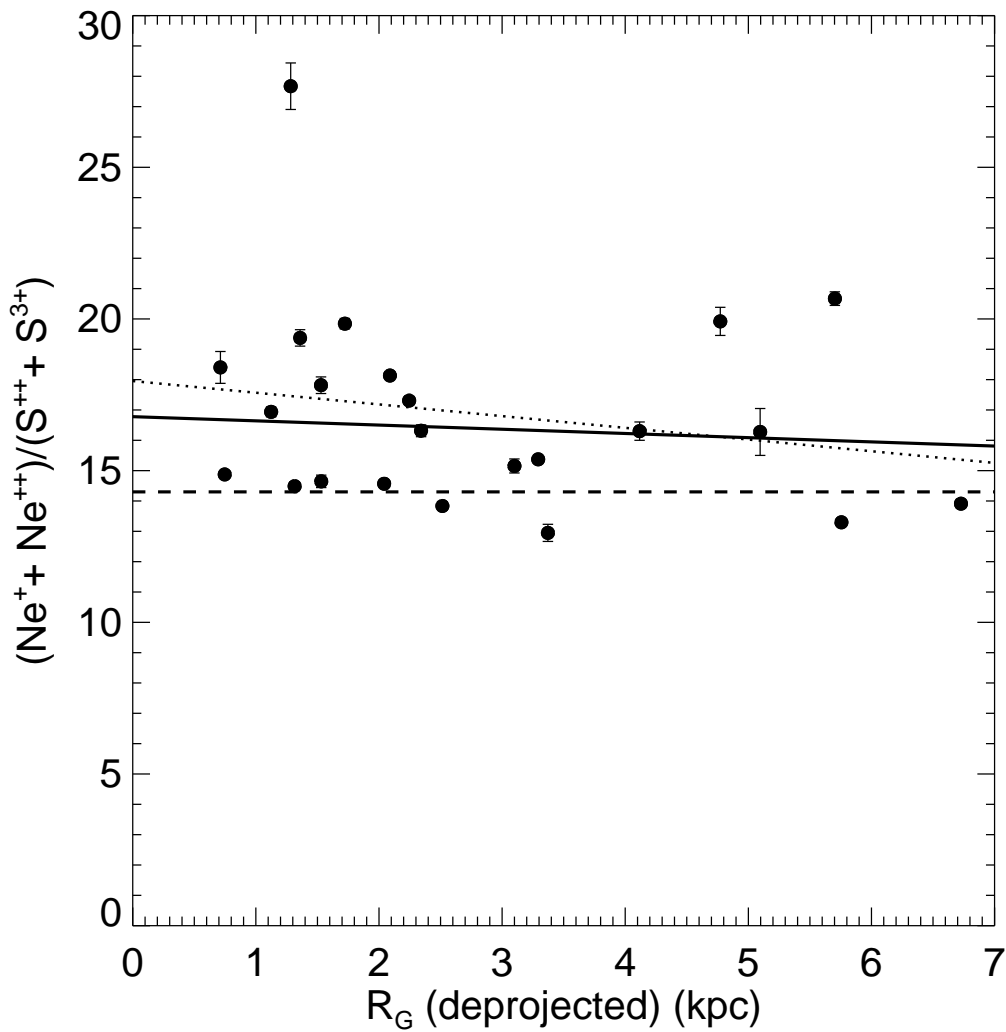} }
\vskip0.1truein
\caption[]{Ne/S, as approximated by
(Ne$^+$ + Ne$^{++}$)/(S$^{++}$ + S$^{3+}$) vs.\ R$_G$.
We fit the points with a linear least-squares function two ways:
using 23 points (omitting source 230 and 702) (dotted line) 
and 22 points, excluding in addition  the most deviant value shown, 
source 32 (solid line).
For both fits, there is no significant variation in the Ne/S ratio with R$_G$.
For  reference, the dashed line depicts a constant Ne/S~= 14.3,
the Orion Nebula value.}

\end{figure}


\begin{figure}
\centering
\resizebox{14.0cm}{!}{\includegraphics{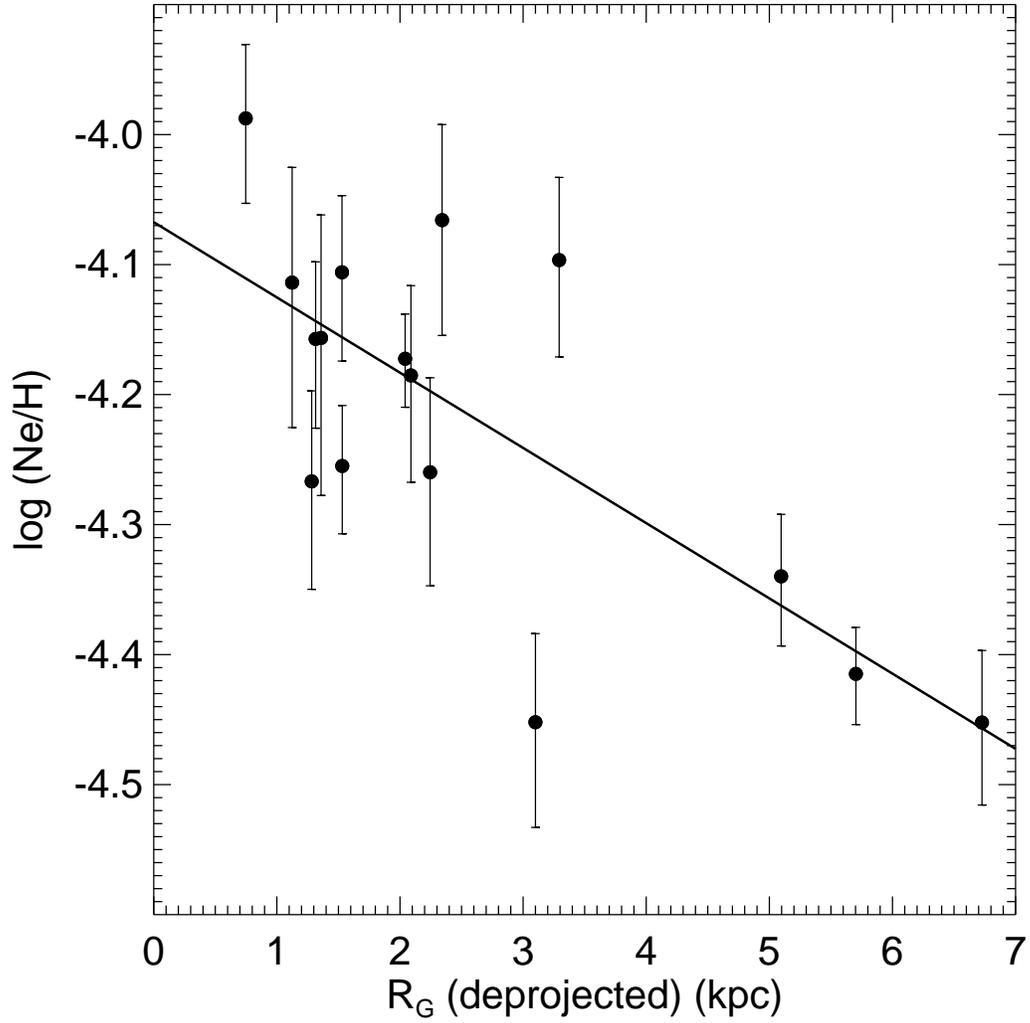} }
\vskip0.1truein
\caption[]{Plot of log~(Ne/H) vs.\ R$_G$ for 16 sources in M33.
The  linear least-squares fit 
is 
log~(Ne/H)~= $-4.07\pm0.04 -(0.058\pm0.014)$~R$_G$.
This Ne/H plot is our most reliable indicator of metallicity
and of a determination of a significant metallicity
gradient with increasing R$_G$.}

\end{figure}


\begin{figure}
\centering
\resizebox{14.0cm}{!}{\includegraphics{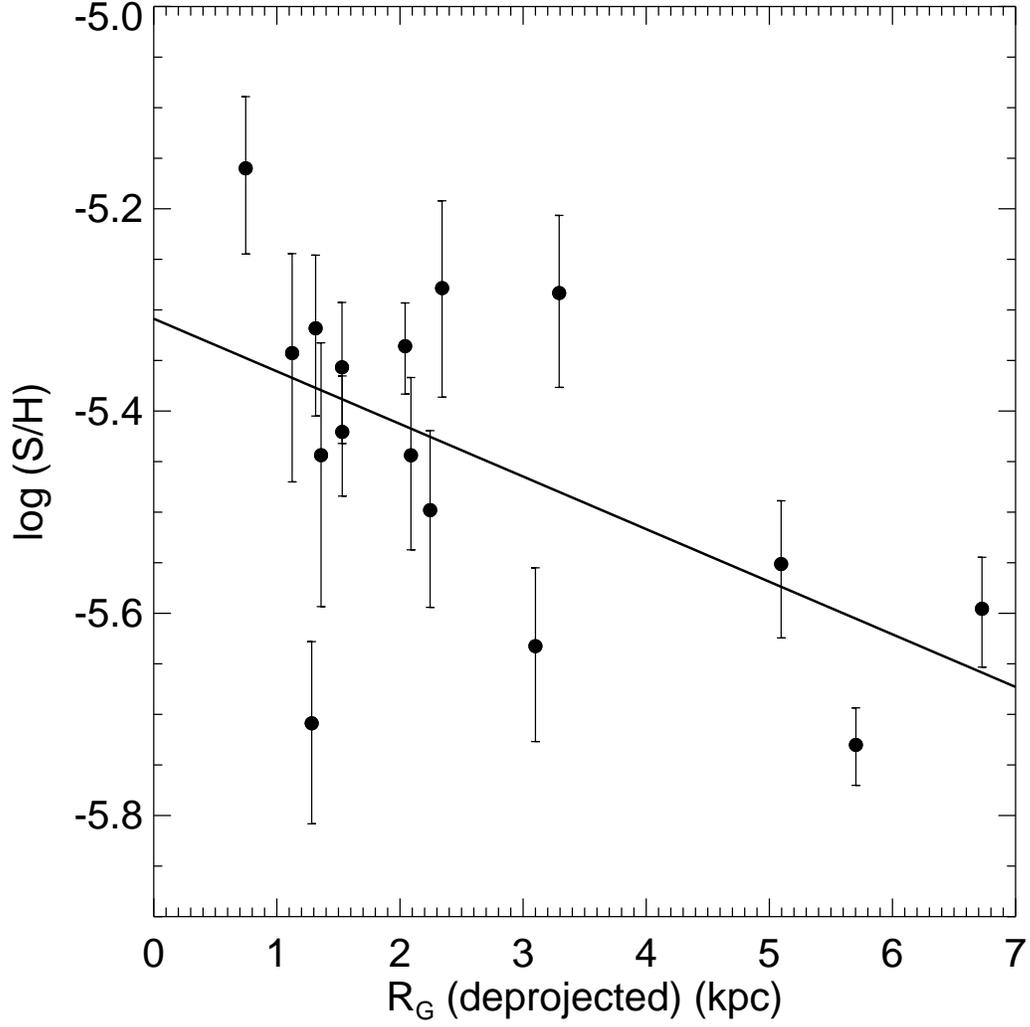} }
\vskip0.1truein
\caption[]{Plot of log~(S/H) vs.\ R$_G$ for 16 sources in M33.
The  linear least-squares fit 
is 
log~(S/H)~= $-5.31\pm0.06 -(0.052\pm0.021)$~R$_G$.
The gradient is similar to that from the Ne/H fit but statistically
less  significant (2.5~$\sigma$).
As discussed in text, this sulfur value does not account for
S$^+$ or for S that may be in dust or molecules.}

\end{figure}


\begin{figure}
  \begin{center}
    \begin{tabular}{cc}
     \resizebox{80mm}{!}{\includegraphics{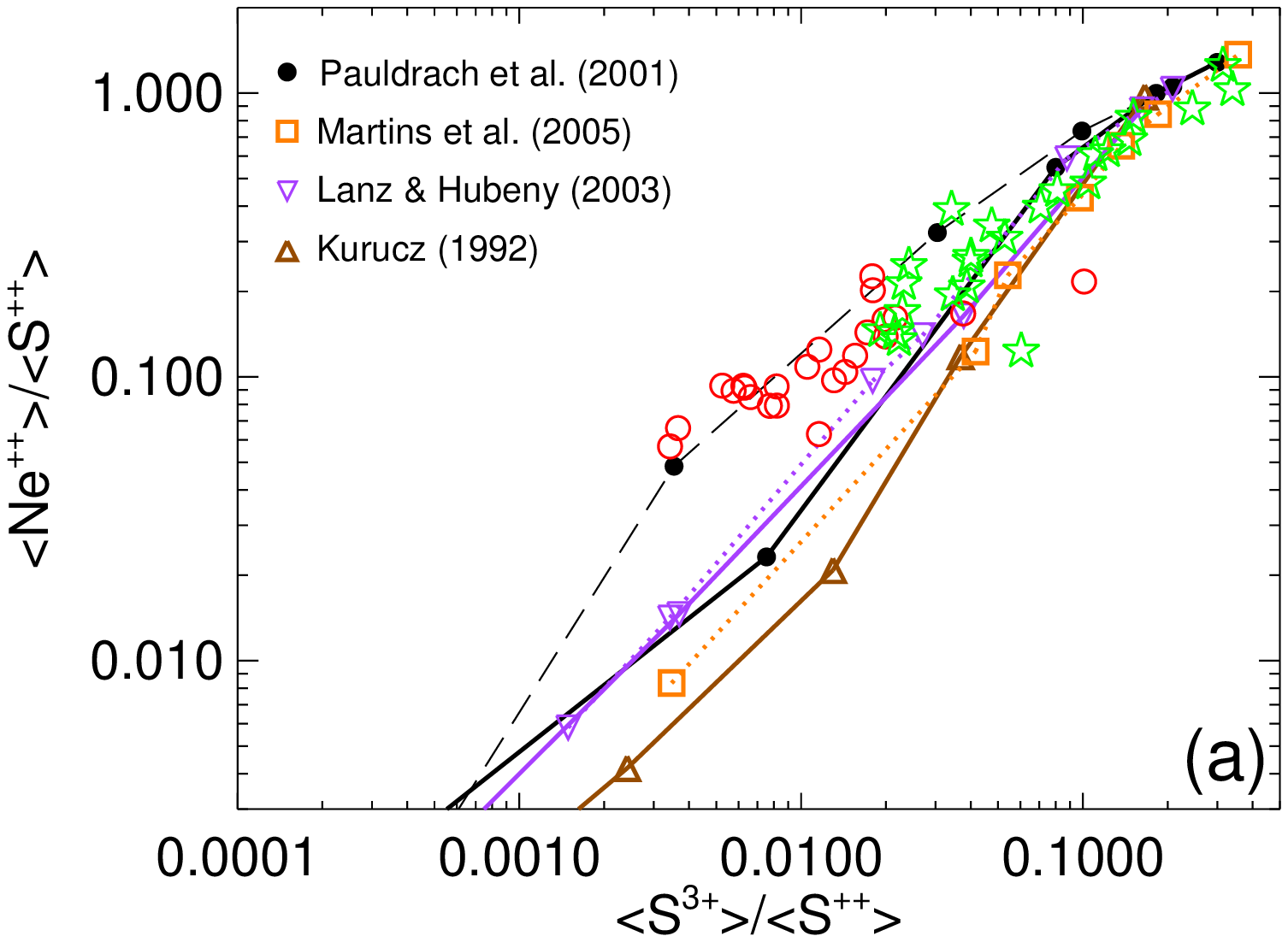}} &
     \resizebox{80mm}{!}{\includegraphics{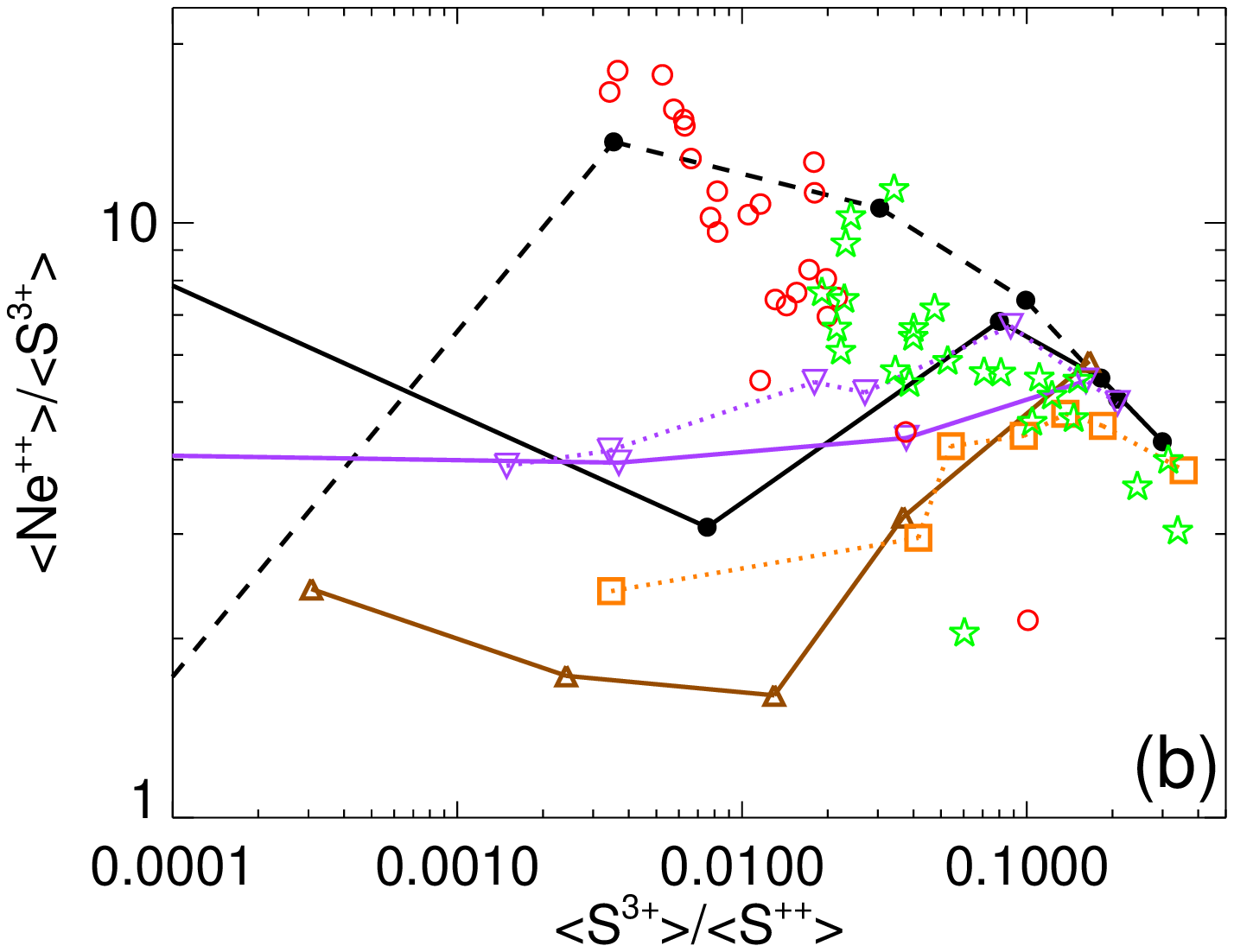}} 
    \end{tabular}
\vskip0.23truein
     \resizebox{80mm}{!}{\includegraphics{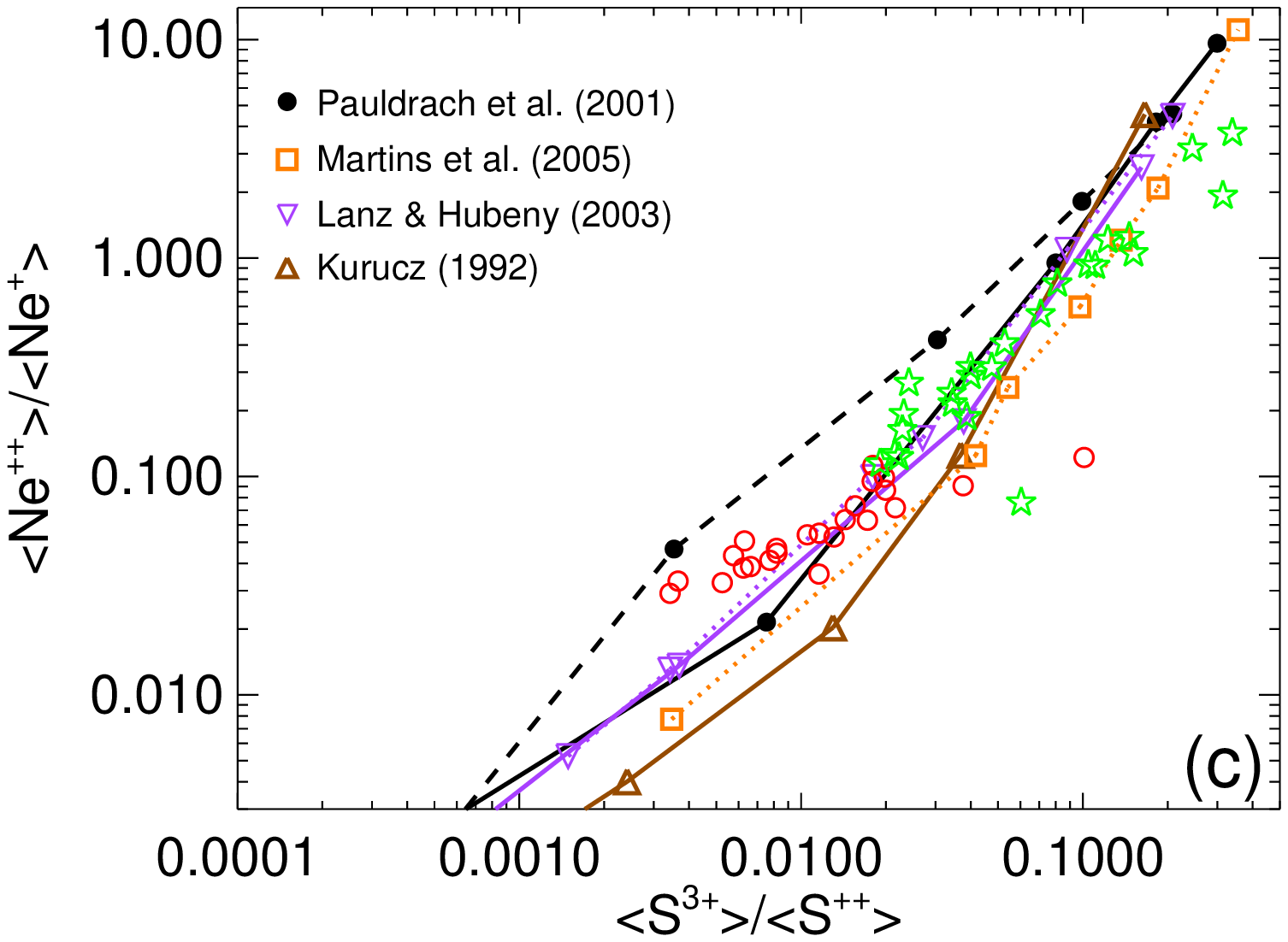}} \\

\caption[]{{\it {\bf (a)}}
Theoretical predictions of the fractional ionization ratios
$<$Ne$^{++}$$>$/$<$S$^{++}$$>$ 
vs.\ 
$<$S$^{3+}$$>$/$<$S$^{++}$$>$,
computed using our photoionization code NEBULA.
The lines connect the results of nebular models calculated with the 
ionizing SEDs predicted from various stellar atmosphere models as labeled,
changing \underbar{no other parameter except the SED}.
For the \ion{H}{2} region models calculated with 
Pauldrach~et~al.\ atmospheres, 
the solid line connects models with dwarf atmospheres and the dashed line 
connects models with supergiant atmospheres.
The violet loci join models calculated with Lanz \& Hubeny atmospheres: 
solid line with log~$g$= 4.0 and the dotted line a set with smaller log~$g$.
The dotted orange line connects models using Martins~et~al.\ atmospheres.
The brown line presents results using Kurucz atmospheres.
The values for \teff\ and log~$g$ for the specific atmospheres used are
described in the text.
To compare our data with the models, we need to divide the observed 
Ne$^{++}$/S$^{++}$
and
Ne$^{++}$/S$^{3+}$
ratios by an assumed Ne/S abundance ratio.
We use the Orion Nebula Ne/S~= 14.3.
The open red circles are our prior results for the M83 \ion{H}{2} regions.
The green stars are the M33 results derived from our observed line fluxes
using $N_e$ of 100~cm$^{-3}$.}

    \label{FIG9}
  \end{center}

\end{figure}
\newpage
\noindent
{\it {\bf (b)}} 
The same as panel (a) except
the ordinate is $<$Ne$^{++}$$>$/$<$S$^{3+}$$>$.
Both panels dramatically
illustrate the sensitivity of the H~{\sc ii}
region model predictions of these ionic abundance ratios to the
ionizing SED that is input to nebular plasma simulations. 
These\break
data, for the most part, appear to track the Pauldrach~et~al.\ 
supergiant locus.\break
{\it {\bf (c)}}
Similar to  panels (a) and (b) except
the ordinate is $<$Ne$^{++}$$>$/$<$Ne$^+$$>$.
This plot also illustrates the sensitivity of the H~{\sc ii}
region model predictions of this ionic abundance ratio to the
ionizing SED that is input to nebular plasma simulations. 
These data, for the most part, appear to 
lie closer to Martins~et~al., Lanz \& Hubeny~et~al., and
Pauldrach~et~al.\ {\it dwarf} loci.


\begin{figure}
  \begin{center}
    \begin{tabular}{cc}
     \resizebox{80mm}{!}{\includegraphics{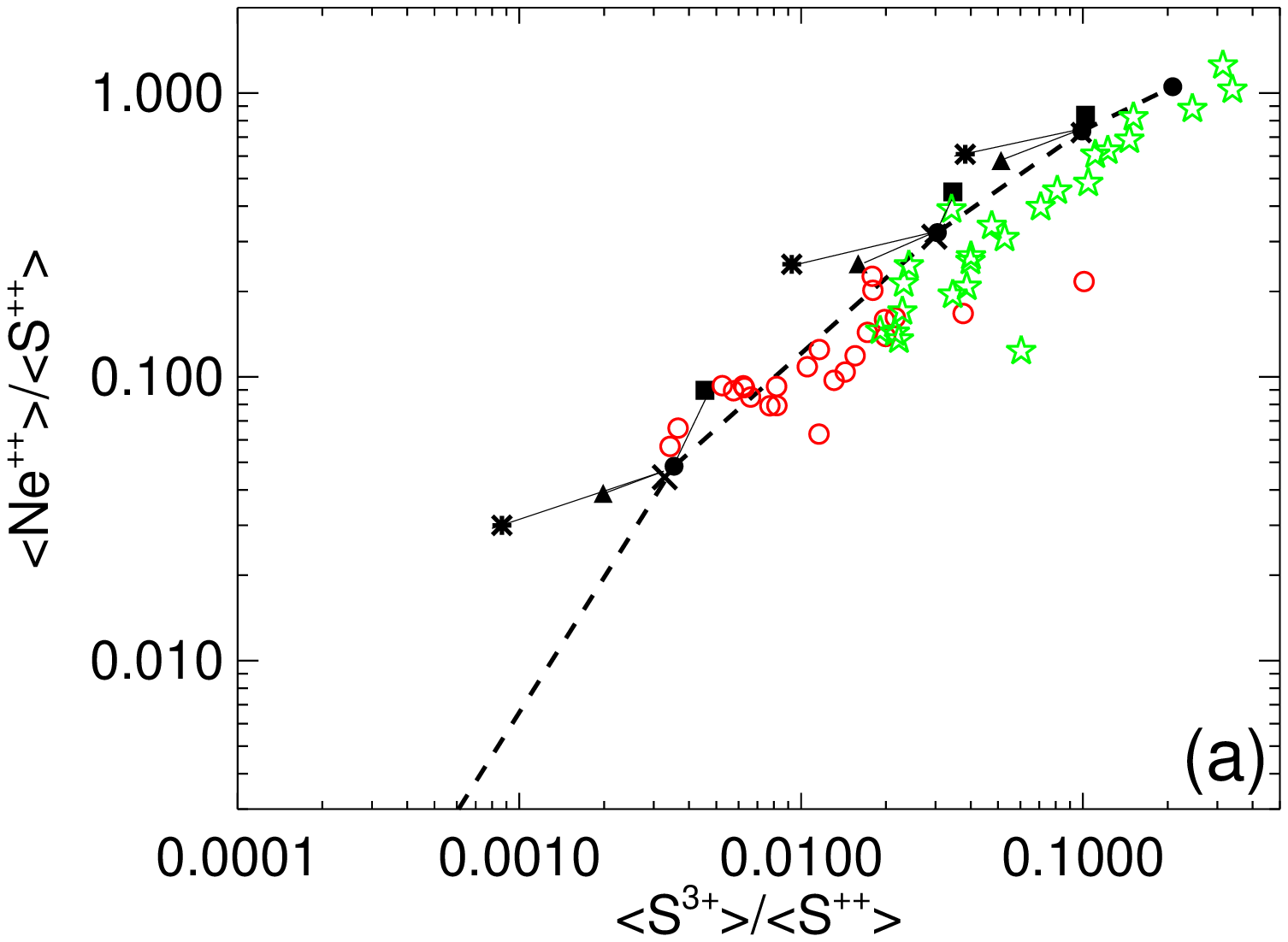}}
     \resizebox{80mm}{!}{\includegraphics{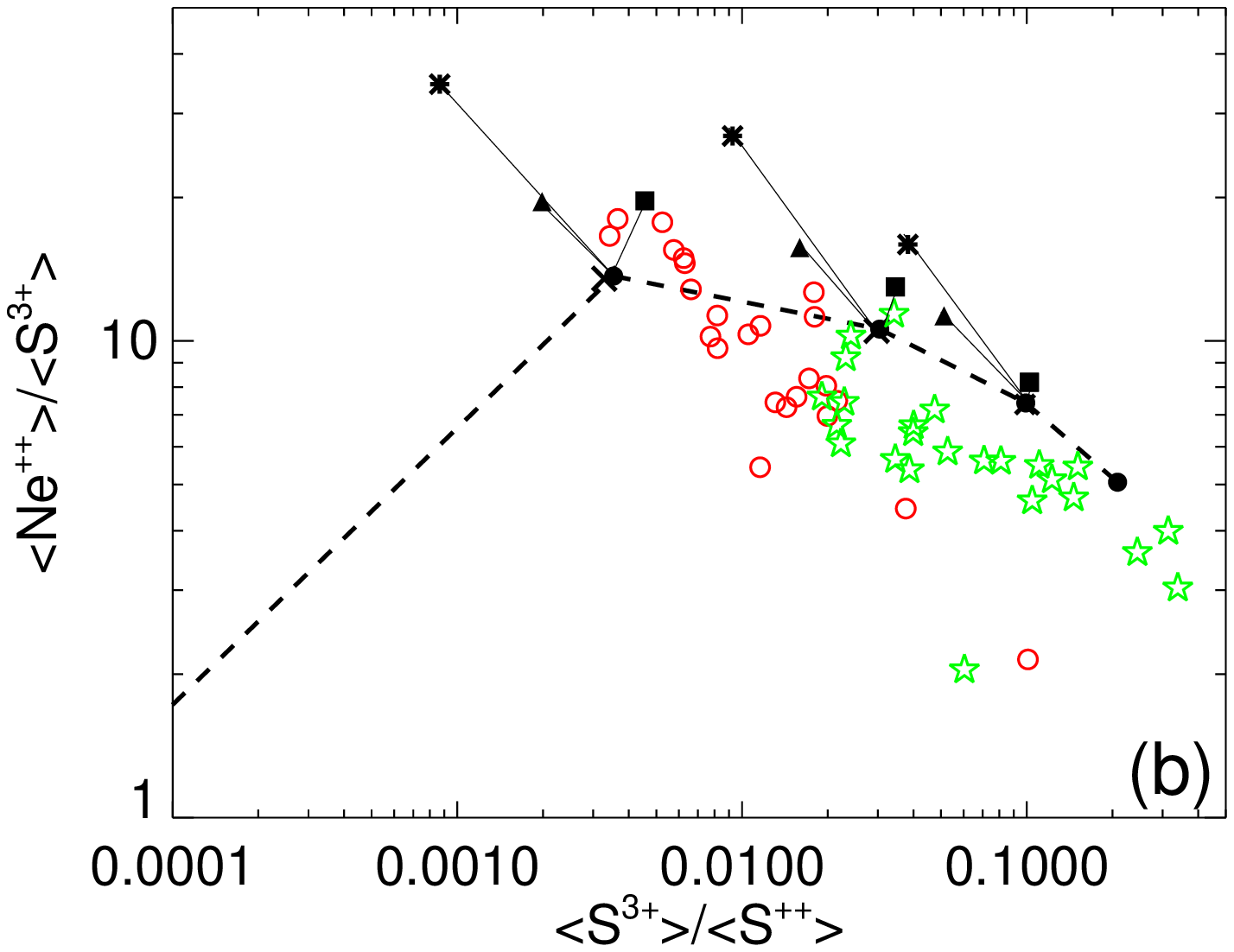}}
    \end{tabular}
\vskip0.23truein
    \begin{tabular}{c}
     \resizebox{80mm}{!}{\includegraphics{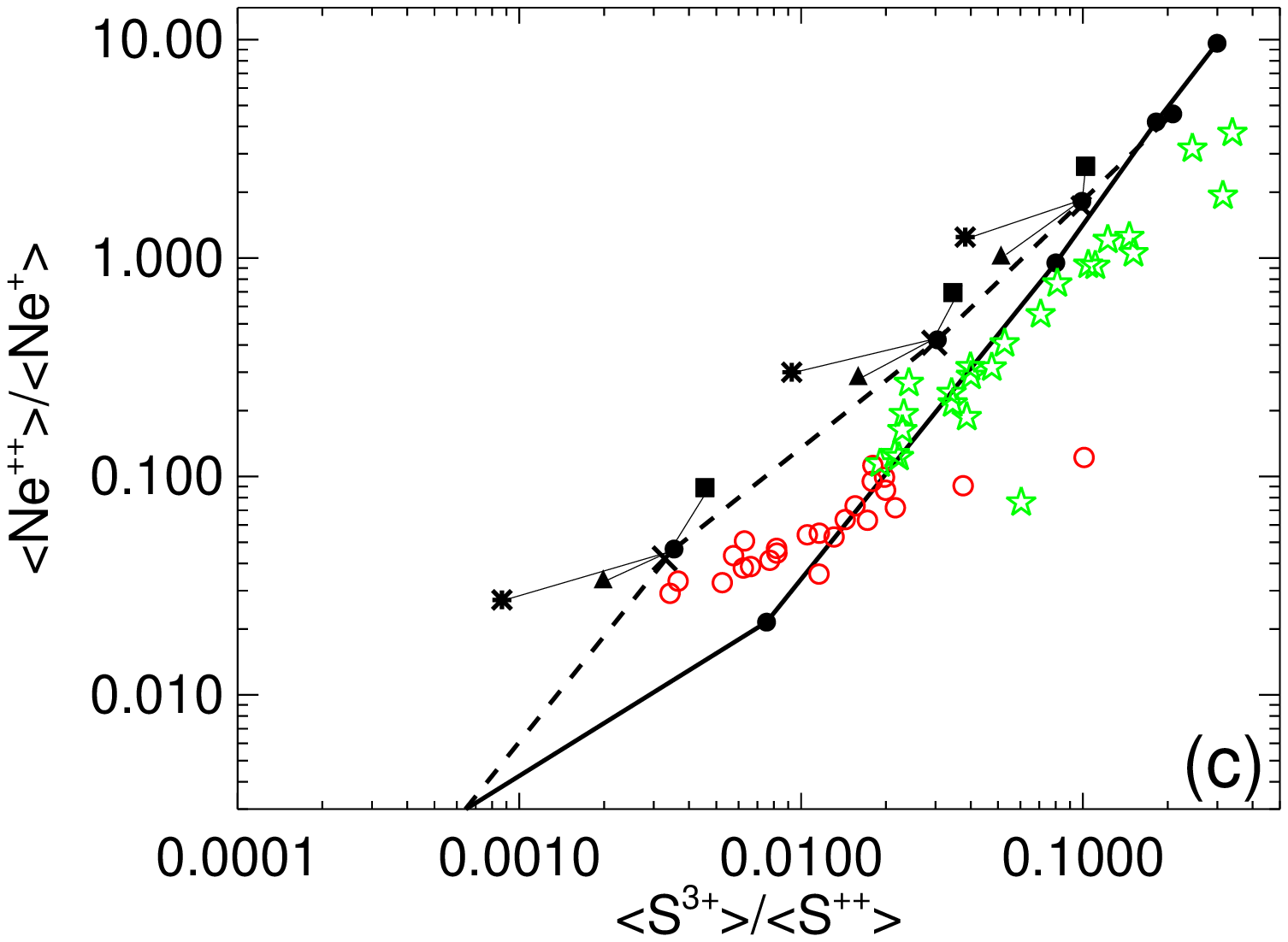}}
    \end{tabular}

\caption[]{{\it {\bf (a)}}
This is similar to Figure~9a.
We again show the locus that uses the 
Pauldrach~et~al.\ (2001) supergiant atmospheres.
Here we display the results of making some changes
to the {\it nebular parameters} for the \teff\ $=$ 35000,
40000, and 45000~K  stellar atmospheres.
The points with an {\bf *} are for a model with a central cavity
of radius 0.5~pc (see text);
those with a triangle have a density of 100 instead
of 1000~cm$^{-3}$;
those with an {\bf X} have a larger number of 
Lyman continuum photons~s$^{-1}$
($N_{Lyc}$)~= 10$^{50}$ instead of 10$^{49}$;
those with a square have all heavy element abundances in  the
nebular set decreased by  a factor  of  three.
The thin solid  lines, emanating  from the standard model points,
trace the change due to each  of  the  four above modifications.
Again, the open circles are from our M83 data while
the open stars are the new M33 results.\break
{\it {\bf (b)}} This is similar to Figure~9b with the same modifications as
described for panel (a).\break
{\it {\bf (c)}} This is similar to Figure~9c with the same modifications as
described for panel (a). Here we also replot the Pauldrach~et~al.\ 
dwarf locus.
A colour version of this figure is available in the electronic edition.}

    \label{FIG10}
  \end{center}

\end{figure}


\begin{figure}
\vskip-0.3truein
\centering
\resizebox{14.0cm}{!}{\includegraphics{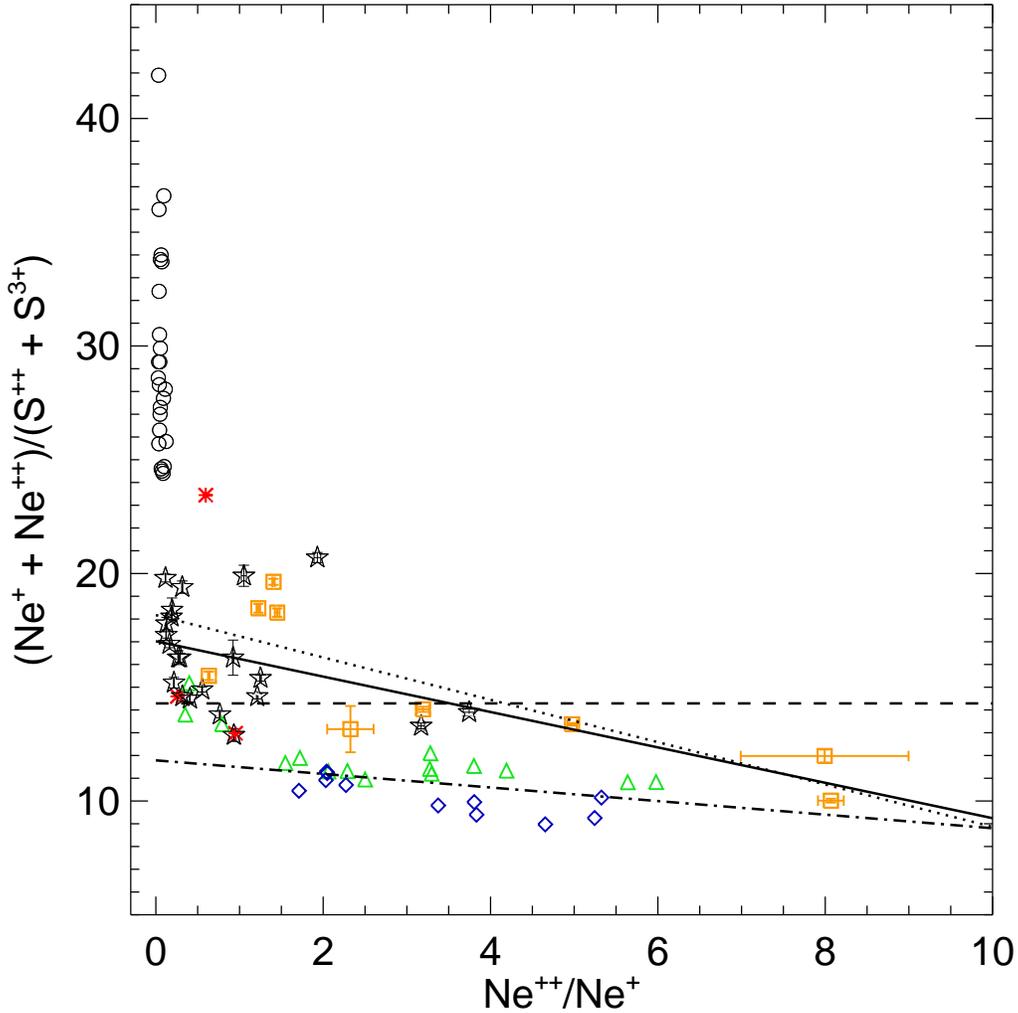} }
\caption[]{Plot of Ne/S vs.\ Ne$^{++}$/Ne$^+$.
Here we show our M33 results as stars for the 22 sources mentioned in  Fig.~6.
The solid line is the linear least-squares fit to the stars
with a gradient that is not statistically significant.
The results from our prior M83 study are shown as circles.
No line fit is done.
These data demonstrate a huge variation in the inferred Ne/S
ratio when Ne$^{++}$/Ne$^+$ is low.
The orange
squares show the 
Wu~et~al.\ (2008) data for 
blue compact dwarf galaxies, 
as reanalyzed with our programme.
We show only 9 points, those objects where they actually detected {\it all 
four} lines: [\ion{S}{4}], [\ion{Ne}{2}], [\ion{Ne}{3}], and [\ion{S}{3}].
The rendition of their data is what we list in columns 2--4 of Table~5.
The dotted line is the linear least-squares fit to the squares
with a statistically significant negative slope (3.8~$\sigma$).
The median (average) Ne/S for the 9 galaxies is 14.0 (14.9), 
very close to  the Orion value of 14.3 shown as the dashed line.
The Lebouteiller~et~al.\ (2008) data were  also reanalyzed here and are
presented as follows: NGC~3603 (red  asterisks), 30~Dor (green triangles),
and N~66 (blue diamonds). 
The median Ne/S ratios for each are 14.6, 11.4, and 10.1, respectively,
possibly indicating a decreasing trend with lower metallicity.
The dash-dot line is a fit to the 23 points in 30~Dor and N~66
that show the highest ionization (see text).}
\end{figure}


\begin{figure}
\centering
\resizebox{14.0cm}{!}{\includegraphics{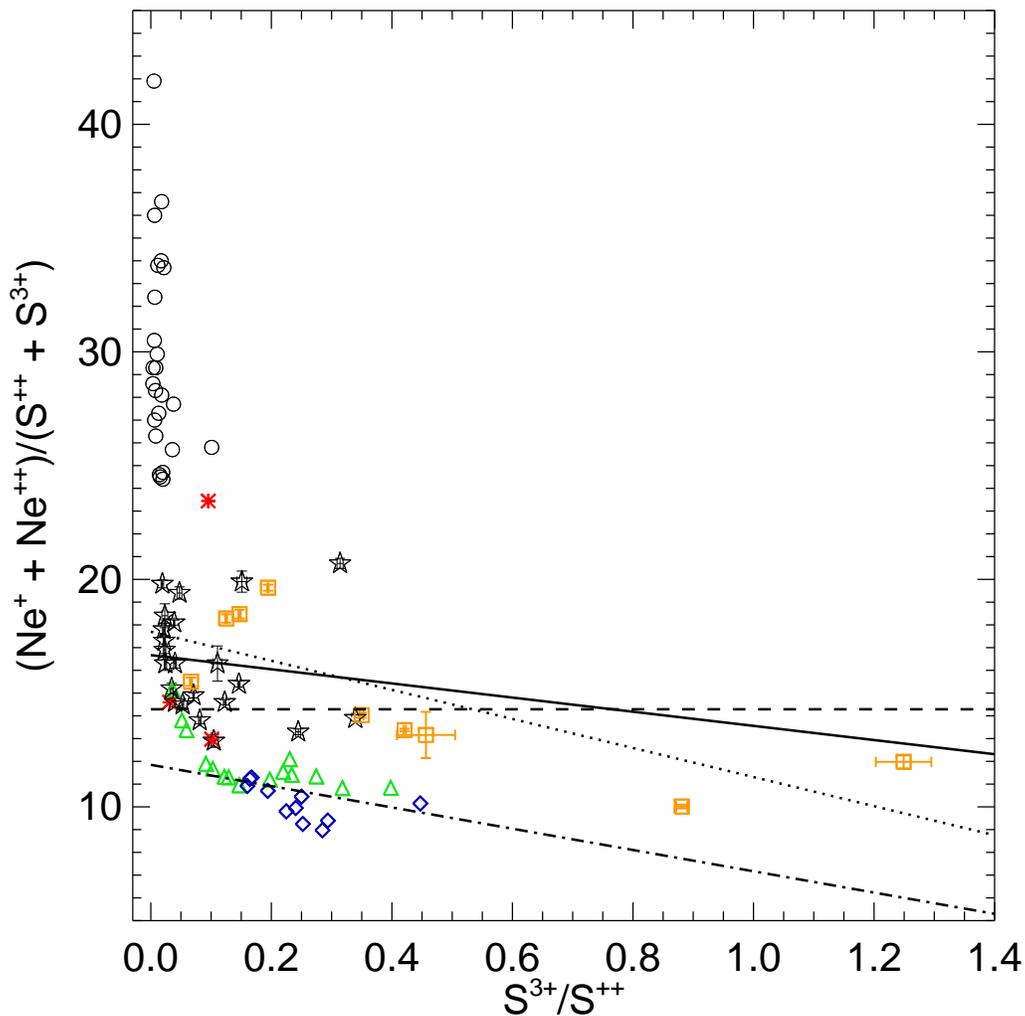} }
\vskip0.1truein
\caption[]{Plot of Ne/S vs.\ S$^{3+}$/S$^{++}$
that is similar to Fig.~11.
Again, the fit to the M33 points results
in a gradient that is not statistically significant.
The M83 points show
a huge variation in the inferred Ne/S
ratio at low ionization, when S$^{3+}$/S$^{++}$ is small.
On the  other  hand, our fit to the Wu~et~al.\ (2008) data produces 
a significant gradient (3.2~$\sigma$).
The least-squares fit to the Lebouteiller~et~al.\ (2008) data 
is for the same 23 points in 30~Dor and N~66 as used in the line fit
in Fig.~11.}

\end{figure}


\begin{deluxetable}{ccccc}
\tabletypesize{\scriptsize}
\tablecaption{H~{\sc ii} Regions Observed in M33}
\tablenum{1}
\tablewidth{0pt}
\tablehead{
\colhead{Order} & \colhead{H~{\sc ii} Region} 
& \colhead{~~~~~~~~~~~~~~~~~RA ~~~~~~J2000} & \colhead{DEC} &
\colhead{Aperture Grid}
}
\startdata
1  & 280 & 1 32 45.4 & 30 38 56 & 2x4 \\

2  & 230 & 1 33 00.7 & 30 34 17 & 1x3 \\

3  & 277 & 1 33 12.2 & 30 38 49 & 2x3 \\
 
4  & 638 & 1 33 16.0 & 30 56 45 & 2x3 \\

5  & 623 & 1 33 16.4 & 30 52 47 & 1x3 \\

6  & 45 & 1 33 29.2 & 30 40 25  & 2x3 \\

7  & 214 & 1 33 30.0 & 30 31 47 & 1x3 \\
 
8  & 33 & 1 33 34.9 & 30 37 06 & 1x3 \\

9  & 42 & 1 33 35.6 & 30 39 30 & 1x2 \\

10 & 32 & 1 33 35.8 & 30 36 29 & 1x3 \\

11 & 251 & 1 33 36.7 & 30 20 13 & 1x3 \\

12 & 62 & 1 33 44.7 & 30 44 38 & 1x3 \\

13 & 27 & 1 33 46.1 & 30 36 54 & 1x2 \\

14 & 301 & 1 33 55.6 & 30 45 27 & 1x3 \\

15 & 4 & 1 33 59.3 & 30 35 48 & 1x3 \\

16 & 79 & 1 34 00.2 & 30 40 51 & 1x3 \\

17 & 87E & 1 34 02.3 & 30 38 45 & 1x3 \\

18 & 302 & 1 34 06.9 & 30 47 27 & 1x3 \\

19 & 702 & 1 34 10.2 & 30 31 54 & 1x2 \\

20 & 95 & 1 34 11.2 & 30 36 16 & 1x3 \\

21 & 710 & 1 34 13.8 & 30 33 44 & 1x3 \\

22 & 88W & 1 34 15.3 & 30 37 11 & 1x3 \\

23 & 691 & 1 34 16.5 & 30 51 56 & 1x3 \\

24 & 651 & 1 34 29.8 & 30 57 15 & 1x3 \\

25 & 740W & 1 34 39.8 & 30 41 54 & 1x3 \\

\enddata

\end{deluxetable}


\begin{deluxetable}{ccccccr}	
\tabletypesize{\scriptsize}
\tablecaption{M33 Line Measurements}
\tablenum{2}
\tablewidth{0pt}
\tablehead{

\colhead{Order} & \colhead{Source} & \colhead{Line} & \colhead{Flux}
& \colhead{1$\sigma$ error} & \colhead{FWHM} & \colhead{V$_{helio}$} \\

& & \colhead{$\mu$m} & \colhead{W cm$^{-2}$} & \colhead{W cm$^{-2}$}

& \colhead{km s$^{-1}$} & \colhead{km s$^{-1}$}

}

\startdata
1  & 280  & 10.5 & 5.71E-20 & 7.76E-22 & 530 & -124 \\
   &      & 12.4 & 1.18E-21$^a$ & 4.28E-22 & 734 & -55 \\
   &      & 12.8 & 1.31E-20 & 1.52E-22 & 496 & -195 \\
   &      & 15.6 & 9.50E-20 & 3.51E-22 & 472 & -147 \\
   &      & 18.7 & 4.96E-20 & 5.24E-22 & 493 & -127 \\

2  & 230  & 10.5 \\
   &      & 12.4 \\
   &      & 12.8 & 6.00E-21 & 1.46E-22 & 578 & -126 \\
   &      & 15.6 & 7.90E-22 & 2.33E-22 & 689 & -68 \\
   &      & 18.7 & 2.90E-21 & 1.90E-22 & 505 & -89 \\

3  & 277  & 10.5 & 1.69E-20 & 9.94E-22 & 556 & -147 \\
   &      & 12.4 \\
   &      & 12.8 & 1.69E-20 & 2.94E-22 & 481 & -207 \\
   &      & 15.6 & 3.61E-20 & 2.32E-22 & 469 & -212 \\
   &      & 18.7 & 3.43E-20 & 7.36E-22 & 588 & -230 \\
 
4  & 638  & 10.5 & 4.94E-20 & 6.96E-22 & 534 & -150 \\
   &      & 12.4 & 1.93E-21 & 2.75E-22 & 588 & -185 \\
   &      & 12.8 & 8.08E-21 & 1.42E-22 & 621 & -208 \\
   &      & 15.6 & 6.92E-20 & 3.02E-22 & 473 & -191 \\
   &      & 18.7 & 3.09E-20 & 3.42E-22 & 491 & -178 \\

5  & 623  & 10.5 & 7.45E-20 & 8.72E-22 & 537 & -203 \\
   &      & 12.4 & 4.21E-21 & 3.26E-22 & 588 & -276 \\
   &      & 12.8 & 3.11E-20 & 6.62E-22 & 479 & -266 \\
   &      & 15.6 & 1.37E-19 & 6.61E-22 & 496 & -225 \\
   &      & 18.7 & 5.04E-20 & 4.66E-22 & 483 & -187 \\

6  & 45   & 10.5 & 1.01E-19 & 1.14E-21 & 533 & -128 \\
   &      & 12.4 & 5.01E-21 & 3.93E-22 & 572 & -161 \\
   &      & 12.8 & 8.57E-20 & 3.65E-22 & 490 & -202 \\
   &      & 15.6 & 2.38E-19 & 8.24E-22 & 481 & -161 \\
   &      & 18.7 & 1.74E-19 & 1.39E-21 & 506 & -141 \\

7  & 214  & 10.5 & 5.42E-21 & 3.25E-22 & 589 & -58 \\
   &      & 12.4 & 1.97E-21 & 3.63E-22 & 743 & -111 \\
   &      & 12.8 & 5.43E-20 & 3.26E-22 & 481 & -135 \\
   &      & 15.6 & 1.53E-20 & 1.75E-22 & 456 & -97 \\
   &      & 18.7 & 5.18E-20 & 3.16E-22 & 519 & -85 \\
 
8  & 33   & 10.5 & 1.22E-20 & 3.51E-22 & 519 & -93 \\
   &      & 12.4 & 1.27E-21 & 2.16E-22 & 513 & -120 \\
   &      & 12.8 & 3.54E-20 & 2.64E-22 & 469 & -192 \\
   &      & 15.6 & 3.30E-20 & 1.37E-22 & 458 & -143 \\
   &      & 18.7 & 4.91E-20 & 4.52E-22 & 508 & -135 \\

9  & 42   & 10.5 & 5.73E-21 & 4.82E-22 & 590 & -19 \\
   &      & 12.4 & 8.81E-22 & 2.58E-22 & 534 & -133 \\
   &      & 12.8 & 2.63E-20 & 2.56E-22 & 501 & -158 \\
   &      & 15.6 & 1.90E-20 & 3.07E-22 & 495 & -89 \\
   &      & 18.7 & 2.56E-20 & 2.93E-22 & 532 & -124 \\

10 & 32   & 10.5 & 2.38E-21 & 2.40E-22 & 592 &  129 \\
   &      & 12.4 & 9.26E-22 & 1.75E-22 & 558 & -100 \\
   &      & 12.8 & 2.27E-20 & 2.19E-22 & 531 & -134 \\
   &      & 15.6 & 1.25E-20 & 1.56E-22 & 495 & -108 \\
   &      & 18.7 & 1.48E-20 & 4.15E-22 & 651 & -135 \\

11 & 251  & 10.5 & 1.57E-20 & 4.39E-22 & 534 & -12 \\
   &      & 12.4 & 1.41E-21 & 1.97E-22 & 439 & -74 \\
   &      & 12.8 & 1.89E-20 & 1.32E-22 & 511 & -86 \\
   &      & 15.6 & 3.99E-20 & 3.12E-22 & 480 & -57 \\
   &      & 18.7 & 3.02E-20 & 1.63E-21 & 526 & -43 \\

12 & 62   & 10.5 & 2.97E-21 & 3.21E-22 & 565 & -123 \\
   &      & 12.4 & 1.71E-21$^a$ & 6.12E-22 & 848 & -479 \\
   &      & 12.8 & 4.00E-20 & 2.48E-22 & 488 & -237 \\
   &      & 15.6 & 1.05E-20 & 1.29E-22 & 473 & -146 \\
   &      & 18.7 & 3.31E-20 & 2.20E-22 & 510 & -173 \\          

13 & 27   & 10.5 & 8.79E-22 & 1.87E-22 & 644 &  37 \\
   &      & 12.4 \\
   &      & 12.8 & 8.47E-21 & 8.19E-23 & 524 & -154 \\
   &      & 15.6 & 3.75E-21 & 1.81E-22 & 499 & -106 \\
   &      & 18.7 & 8.07E-21 & 2.23E-22 & 514 & -90 \\

14 & 301  & 10.5 & 4.75E-21 & 4.39E-22 & 639 & -132 \\
   &      & 12.4 & 1.29E-21 & 1.78E-22 & 471 & -210 \\
   &      & 12.8 & 5.04E-20 & 5.64E-22 & 476 & -226 \\
   &      & 15.6 & 1.46E-20 & 2.55E-22 & 480 & -167 \\
   &      & 18.7 & 4.69E-20 & 5.63E-22 & 493 & -172 \\

15 & 4    & 10.5 & 1.46E-20 & 4.34E-22 & 524 & -159 \\
   &      & 12.4 & 2.52E-21 & 2.84E-22 & 778 & -258 \\
   &      & 12.8 & 6.00E-20 & 2.32E-22 & 466 & -240 \\
   &      & 15.6 & 4.35E-20 & 2.15E-21 & 463 & -200 \\
   &      & 18.7 & 7.80E-20 & 5.68E-22 & 507 & -190 \\

16 & 79   & 10.5 & 7.70E-20 & 1.26E-21 & 568 & -226 \\
   &      & 12.4 & 4.23E-21 & 7.16E-22 & 427 & -283 \\
   &      & 12.8 & 1.58E-19 & 1.24E-21 & 449 & -331 \\
   &      & 15.6 & 1.99E-19 & 6.83E-22 & 489 & -221 \\
   &      & 18.7 & 2.31E-19 & 1.40E-21 & 508 & -236 \\

17 & 87E  & 10.5 & 5.29E-21 & 4.49E-22 & 582 & -166 \\
   &      & 12.4 & 1.31E-21 & 3.21E-22 & 638 & -392 \\
   &      & 12.8 & 4.87E-20 & 2.81E-22 & 449 & -148 \\
   &      & 15.6 & 1.82E-20 & 2.51E-22 & 463 & -193 \\
   &      & 18.7 & 4.91E-20 & 4.06E-22 & 507 & -184 \\

18 & 302  & 10.5 & 7.26E-21 & 1.76E-22 & 410 & -227 \\
   &      & 12.4 & 1.36E-21 & 2.47E-22 & 565 & -326 \\
   &      & 12.8 & 4.21E-20 & 1.93E-22 & 438 & -299 \\
   &      & 15.6 & 1.80E-20 & 1.29E-22 & 464 & -232 \\
   &      & 18.7 & 3.98E-20 & 1.92E-22 & 496 & -246 \\

19 & 702  & 10.5 & 8.60E-22 & 2.62E-22 & 533 &  374 \\
   &      & 12.4 \\
   &      & 12.8 & 4.67E-21 & 1.62E-22 & 576 & -104 \\
   &      & 15.6 & 8.13E-22 & 2.00E-22 & 903 & -69  \\
   &      & 18.7 & 3.03E-21 & 1.86E-22 & 574 & -95  \\

20 & 95   & 10.5 & 6.95E-21 & 3.86E-22 & 595 & -118 \\
   &      & 12.4 & 8.60E-22 & 1.82E-22 & 462 & -236 \\
   &      & 12.8 & 3.23E-20 & 2.35E-22 & 473 & -215 \\
   &      & 15.6 & 2.13E-20 & 2.56E-22 & 471 & -163 \\
   &      & 18.7 & 3.68E-20 & 3.98E-22 & 532 & -161 \\

21 & 710  & 10.5 & 6.59E-21 & 5.02E-22 & 647 & -66 \\
   &      & 12.4 & 2.13E-21 & 3.88E-22 & 893 & -84 \\
   &      & 12.8 & 3.48E-20 & 4.07E-22 & 515 & -155 \\
   &      & 15.6 & 1.72E-20 & 2.00E-22 & 522 & -108 \\
   &      & 18.7 & 4.05E-20 & 4.90E-22 & 554 & -119 \\

22 & 88W  & 10.5 & 1.55E-20 & 3.34E-22 & 548 & -143 \\
   &      & 12.4 \\
   &      & 12.8 & 2.29E-20 & 1.26E-22 & 444 & -196 \\
   &      & 15.6 & 4.01E-20 & 2.46E-22 & 470 & -163 \\
   &      & 18.7 & 4.06E-20 & 4.21E-22 & 521 & -159 \\

23 & 691  & 10.5 & 4.43E-20 & 5.36E-22 & 509 & -238 \\
   &      & 12.4 & 1.67E-21 & 3.42E-22 & 655 & -406 \\ 
   &      & 12.8 & 3.35E-20 & 2.99E-22 & 489 & -286 \\
   &      & 15.6 & 9.60E-20 & 4.39E-22 & 488 & -242 \\
   &      & 18.7 & 6.43E-20 & 5.49E-22 & 504 & -226 \\
    
24 & 651  & 10.5 & 5.67E-21 & 2.63E-22 & 541 & -195 \\
   &      & 12.4 \\
   &      & 12.8 & 5.93E-21 & 1.67E-22 & 575 & -276 \\
   &      & 15.6 & 1.43E-20 & 1.20E-22 & 509 & -199 \\
   &      & 18.7 & 7.97E-21 & 1.63E-22 & 544 & -258 \\

25 & 740W & 10.5 & 2.29E-21 & 1.90E-22 & 375 & -146 \\ 
   &      & 12.4 \\
   &      & 12.8 & 1.77E-20 & 2.60E-22 & 482 & -224 \\
   &      & 15.6 & 1.09E-20 & 1.58E-22 & 468 & -179 \\
   &      & 18.7 & 2.02E-20 & 2.99E-22 & 504 & -164 \\

\enddata

$^a$ Flux less than 3$\sigma$, not used in the analysis

\end{deluxetable}

\begin{deluxetable}{rrccccccccccccc} 
\rotate
\setlength{\tabcolsep}{0.04in}
\tabletypesize{\scriptsize}
\tablecaption{Derived Parameters for the H~{\sc ii} Regions in M33}
\tablenum{3}
\tablewidth{0pt}
\tablehead{
\\ 
& \colhead{Source}
& \colhead{R$_G$}
& \colhead{{\underbar{Ne$^+$}}}
& \colhead{\underbar{Ne$^{++}$}}
& \colhead{\underbar{S$^{++}$}}
& \colhead{\underbar{S$^{3+}$}}
& \colhead{{\underbar{Ne$^+$}}}
& \colhead{\underbar{Ne$^{++}$}}
& \colhead{\underbar{Ne$^{++}$}}
& \colhead{\underbar{S$^{3+}$}}
& \colhead{\underbar{Ne}}
& \colhead{\underbar{$<$S$^{++}$$>$}}
& \colhead{\underbar{$<$Ne$^{++}$$>$}}
& \colhead{\underbar{$<$Ne$^{++}$$>$}}

\\ 
& 
& \colhead{kpc}
& \colhead{H$^+$}
& \colhead{H$^+$}
& \colhead{H$^+$}
& \colhead{H$^+$}
& \colhead{S$^{++}$}
& \colhead{S$^{++}$}
& \colhead{Ne$^+$}
& \colhead{S$^{++}$}
& \colhead{S}
& \colhead{$<$Ne$^+$$>$}
& \colhead{$<$S$^{++}$$>$}
& \colhead{$<$S$^{3+}$$>$}
\\ 
& 
& 
& \colhead{$($$\times${10$^{-6}$}$)$}
& \colhead{$($$\times${10$^{-6}$}$)$}
& \colhead{$($$\times${10$^{-6}$}$)$}
& \colhead{$($$\times${10$^{-8}$}$)$}
& 
& 
& 
& 
& 
& 
& 
& 
}

\startdata

1 & 280 & 5.76 &  &  &  &  & 3.97$\pm$0.06 & 12.6$\pm$0.1 & 3.17$\pm$0.04 & 0.245$\pm$0.004 & 13.3$\pm$0.1 & 3.60$\pm$0.06 & 0.881$\pm$0.010 & 3.60$\pm$0.05 \\

2 & 230 & 4.11 &  &  &  &  & 31.0$\pm$2.1 & 1.79$\pm$0.54 & 0.0575$\pm$0.0170 & & 32.8$\pm$2.3 & 0.460$\pm$0.031 & 0.125$\pm$0.038 \\

3 & 277 & 3.37 &  &  &  &  & 7.40$\pm$0.20 & 6.90$\pm$0.15 & 0.932$\pm$0.017 & 0.104$\pm$0.007 & 12.9$\pm$0.3 & 1.93$\pm$0.05 & 0.483$\pm$0.010 & 4.63$\pm$0.27 \\

4 & 638 & 6.73 & 8.35$\pm$1.40 & 31.3$\pm$5.2 & 2.13$\pm$0.35 & 72.1$\pm$12.0 & 3.93$\pm$0.08 & 14.7$\pm$0.2 & 3.74$\pm$0.07 & 0.339$\pm$0.006 & 13.9$\pm$0.1 & 3.64$\pm$0.08 & 1.03$\pm$0.01 & 3.04$\pm$0.04 \\

5 & 623 & 5.70 & 14.7$\pm$1.7 & 28.4$\pm$3.3 & 1.59$\pm$0.18 & 49.9$\pm$5.8 & 9.27$\pm$0.21 & 17.9$\pm$0.2 & 1.93$\pm$0.04 & 0.314$\pm$0.005 & 20.7$\pm$0.2 & 1.54$\pm$0.04 & 1.25$\pm$0.01 & 3.99$\pm$0.05 \\

6 & 45 & 2.04 & 34.1$\pm$4.0 & 41.3$\pm$4.8 & 4.61$\pm$0.54 & 56.5$\pm$6.6 & 7.39$\pm$0.07 & 8.96$\pm$0.08 & 1.21$\pm$0.01 & 0.123$\pm$0.002 & 14.6$\pm$0.1 & 1.93$\pm$0.02 & 0.627$\pm$0.005 & 5.12$\pm$0.06 \\

7 & 214 & 2.25 & 55.0$\pm$11.2 & 6.75$\pm$1.37 & 3.49$\pm$0.71 & 7.75$\pm$1.64 & 15.8$\pm$0.1 & 1.93$\pm$0.02 & 0.123$\pm$0.002 & 0.0222$\pm$0.0013 & 17.3$\pm$0.1 & 0.906$\pm$0.008 & 0.135$\pm$0.002 & 6.09$\pm$0.37 \\

8 & 33 & 1.32 & 55.5$\pm$10.6 & 22.6$\pm$4.3 & 5.12$\pm$0.98 & 27.0$\pm$5.2 & 10.8$\pm$0.1 & 4.41$\pm$0.04 & 0.407$\pm$0.003 & 0.0527$\pm$0.0016 & 14.5$\pm$0.1 & 1.32$\pm$0.02 & 0.309$\pm$0.003 & 5.86$\pm$0.17 \\

9 & 42 & 1.36 & 59.5$\pm$18.2 & 18.8$\pm$5.7 & 3.86$\pm$1.18 & 18.3$\pm5.8$ & 15.4$\pm$0.2 & 4.87$\pm$0.10 & 0.315$\pm$0.006 & 0.0475$\pm$0.0040 & 19.4$\pm$0.3 & 0.925$\pm$0.014 & 0.341$\pm$0.007 & 7.18$\pm$0.62 \\

10 & 32 & 1.28 & 48.9$\pm$10.3 & 11.8$\pm$2.5 & 2.12$\pm$0.45 & 7.25$\pm$1.68 & 23.1$\pm$0.7 & 5.56$\pm$0.17 & 0.241$\pm$0.004 & 0.0342$\pm$0.0036 & 27.7$\pm$0.8 & 0.619$\pm$0.018 & 0.390$\pm$0.012 & 11.4$\pm$1.2 \\

11 & 251 & 5.10 & 26.7$\pm$4.4 & 24.6$\pm$4.0 & 2.84$\pm$0.49 & 31.4$\pm$5.2 & 9.40$\pm$0.50 & 8.67$\pm$0.46 & 0.922$\pm$0.010 & 0.111$\pm$0.007 & 16.3$\pm$0.8 & 1.52$\pm$0.08 & 0.607$\pm$0.032 & 5.49$\pm$0.16 \\

12 & 62 & 1.72 &  &  &  &  & 18.1$\pm$0.2 & 2.08$\pm$0.03 & 0.115$\pm$0.002 & 0.0191$\pm$0.0021 & 19.8$\pm$0.2 & 0.787$\pm$0.007 & 0.146$\pm$0.002 & 7.68$\pm$0.83 \\

13 & 27 & 0.712 &  &  &  &  & 15.8$\pm$0.4 & 3.05$\pm$0.17 & 0.193$\pm$0.010 & 0.0231$\pm$0.0050 & 18.4$\pm$0.5 & 0.905$\pm$0.026 & 0.214$\pm$0.012 & 9.23$\pm$2.01\\

14 & 301 & 1.53 & 78.0$\pm$12.7 & 9.89$\pm$1.61 & 4.83$\pm$0.79 & 10.4$\pm$1.9 & 16.2$\pm$0.3 & 2.05$\pm$0.04 & 0.127$\pm$0.003 & 0.0215$\pm$0.0020 & 17.8$\pm$0.3 & 0.884$\pm$0.014 & 0.143$\pm$0.003 & 6.66$\pm$0.63 \\

15 & 4 & 1.53 & 47.4$\pm$6.7 & 15.0$\pm$2.2 & 4.10$\pm$0.58 & 16.4$\pm$2.4 & 11.6$\pm$0.1 & 3.66$\pm$0.18 & 0.317$\pm$0.016 & 0.0399$\pm$0.0012 & 14.6$\pm$0.2 & 1.23$\pm$0.01 & 0.257$\pm$0.013 & 6.43$\pm$0.37 \\

16 & 79 & 0.747 & 74.4$\pm$14.1 & 41.1$\pm$7.8 & 7.25$\pm$1.38 & 51.3$\pm$9.8 & 10.3$\pm$0.1 & 5.67$\pm$0.04 & 0.553$\pm$0.005 & 0.0708$\pm$0.0012 & 14.9$\pm$0.1 & 1.39$\pm$0.01 & 0.397$\pm$0.003 & 5.61$\pm$0.09 \\

17 & 87E & 1.12 & 74.2$\pm$19.3 & 12.1$\pm$3.2 & 4.98$\pm$1.30 & 11.4$\pm$3.1 & 14.9$\pm$0.1 & 2.43$\pm$0.04 & 0.163$\pm$0.002 & 0.0229$\pm$0.0019 & 16.9$\pm$0.2 & 0.959$\pm$0.009 & 0.170$\pm$0.003 & 7.44$\pm$0.64 \\

18 & 302 & 2.09 & 61.7$\pm$12.4 & 11.6$\pm$2.3 & 3.89$\pm$0.78 & 15.1$\pm$3.1 & 15.9$\pm$0.1 & 2.97$\pm$0.03 & 0.187$\pm$0.002 & 0.0387$\pm$0.0010 & 18.1$\pm$0.1 & 0.900$\pm$0.006 & 0.208$\pm$0.002 & 5.37$\pm$0.14 \\

19 & 702 & 3.27 &  &  &  &  & 23.2$\pm$1.6 & 1.76$\pm$0.45 & 0.0761$\pm$0.0189 & 0.0604$\pm$0.0188 & 23.5$\pm$1.6 & 0.616$\pm$0.042 & 0.123$\pm$0.031 & 2.04$\pm$0.80 \\

20 & 95 & 2.34 & 74.8$\pm$17.1 & 21.6$\pm$5.0 & 5.68$\pm$1.30 & 22.8$\pm$5.4 & 13.2$\pm$0.2 & 3.80$\pm$0.06 & 0.289$\pm$0.004 & 0.0401$\pm$0.0023 & 16.3$\pm$0.2 & 1.08$\pm$0.01 & 0.266$\pm$0.004 & 6.65$\pm$0.38 \\

21 & 710 & 3.10 & 32.6$\pm$6.6 & 7.04$\pm$1.42 & 2.53$\pm$0.51 & 8.73$\pm$1.88 & 12.9$\pm$0.2 & 2.78$\pm$0.05 & 0.216$\pm$0.004 & 0.0345$\pm$0.0027 & 15.2$\pm$0.2 & 1.11$\pm$0.02 & 0.195$\pm$0.003 & 5.65$\pm$0.44 \\

22 & 88W & 2.52 &  &  &  &  & 8.48$\pm$0.10 & 6.48$\pm$0.08 & 0.765$\pm$0.006 & 0.0811$\pm$0.0019 & 13.8$\pm$0.1 & 1.68$\pm$0.02 & 0.454$\pm$0.005 & 5.60$\pm$0.13 \\

23 & 651 & 4.77 &  &  &  &  & 11.2$\pm$0.4 & 11.8$\pm$0.3 & 1.05$\pm$0.03 & 0.151$\pm$0.008 & 19.9$\pm$0.5 & 1.28$\pm$0.04 & 0.823$\pm$0.018 & 5.45$\pm$0.26 \\

24 & 691 & 3.29 & 39.9$\pm$8.8 & 49.9$\pm$11.1 & 5.10$\pm$1.13 & 74.5$\pm$16.5 & 7.83$\pm$0.10 & 9.79$\pm$0.09 & 1.25$\pm$0.01 & 0.146$\pm$0.002 & 15.4$\pm$0.1 & 1.82$\pm$0.02 & 0.686$\pm$0.006 & 4.69$\pm$0.06 \\

25 & 740W & 4.12 &  &  &  &  & 13.2$\pm$0.3 & 3.53$\pm$0.07 & 0.268$\pm$0.006 & 0.0241$\pm$0.0020 & 16.3$\pm$0.3 & 1.08$\pm$0.02 & 0.247$\pm$0.005 & 10.3$\pm$0.9 \\
\enddata
\end{deluxetable}


\def\Term#1 #2 #3/{\mbox{$\,^{#1}\!#2_{#3}$ }}
\def\Termo#1 #2 #3/{\mbox{$\,^{#1}\!#2^o_{#3}$ }}
\def\sterm #1 #2 #3/{\mbox{$\,_{#3}\!^{#1}\!#2$}}

\newcommand{\tef}{$T_{\rm eff}$}
\newcommand{\rinit}{$R_{\rm init}$}
\newcommand{\nlyc}{$N_{\rm Lyc}$}

\begin{deluxetable}{rccrccc}	
\tabletypesize{\scriptsize}
\tablecaption{Models Varying the Nebular Parameters}
\tablewidth{0pt}
\tablenum{4}
\tablehead{
  \colhead{Model}
& \colhead{Symbol} 
& \colhead{\tef}
& \colhead{DENS}
& \colhead{\rinit}
& \colhead{\nlyc}
& \colhead{Abundances}

\\ 
& 
& \colhead{(K)}
& \colhead{(cm$^{-3}$)}
& \colhead{(pc)}
& \colhead{(s$^{-1}$)}
& 

}

\startdata


1  & asterisk & 35000  & 1000 & 0.5 & $10^{49}$ & reference \\

2  & triangle & 35000  & 100 & 0.0 & $10^{49}$ & reference \\

3  & X & 35000  & 100 & 0.0 & $10^{50}$ & reference \\

4  & square & 35000  & 1000 & 0.0 & $10^{49}$ & reference/3 \\

5  & asterisk & 40000  & 1000 & 0.5 & $10^{49}$ & reference \\

6  & triangle & 40000  & 100 & 0.0 & $10^{49}$ & reference \\

7  & X & 40000  & 100 & 0.0 & $10^{50}$ & reference \\

8  & square & 40000  & 1000 & 0.0 & $10^{49}$ & reference/3 \\

9  & asterisk & 45000  & 1000 & 0.5 & $10^{49}$ & reference \\

10 & triangle & 45000  & 100 & 0.0 & $10^{49}$ & reference \\

11 & X & 45000  & 100 & 0.0 & $10^{50}$ & reference \\

12 & square & 45000  & 1000 & 0.0 & $10^{49}$ & reference/3 \\

\enddata

\end{deluxetable}


\def\Term#1 #2 #3/{\mbox{$\,^{#1}\!#2_{#3}$ }}
\def\Termo#1 #2 #3/{\mbox{$\,^{#1}\!#2^o_{#3}$ }}
\def\sterm #1 #2 #3/{\mbox{$\,_{#3}\!^{#1}\!#2$}}

\newcommand{\te}{$T_e$}
\newcommand{\nelec}{$N_e$}

\begin{deluxetable}{ccccccccc}	
\tabletypesize{\scriptsize}
\tablecaption{Blue Compact Dwarf Galaxy Results}
\tablewidth{0pt}
\setlength{\tabcolsep}{0.06in}
\tablenum{5}
\tablehead{

\colhead{} & \multicolumn{3}{c}{\te~\&~\nelec~from Wu et al. (2008) Table 2}
& \colhead{} & \colhead{} & \multicolumn{3}{c}{\te~=~10000~K \&~\nelec~=~100 cm$^{-3}$} \\
\cline{2-4} \cline{7-9}

\colhead{Object} & \colhead{Ne$^{++}$/Ne$^+$} & \colhead{S$^{3+}$/S$^{++}$}
& \colhead{Ne/S} & \colhead{} & \colhead{}  & \colhead{Ne$^{++}$/Ne$^+$} 
& \colhead{S$^{3+}$/S$^{++}$} & \colhead{Ne/S}

}

\startdata


Haro11 & 1.41$\pm$0.01 & 0.194$\pm$0.001 & 19.6$\pm$0.1 & & & 1.38$\pm$0.01 & 0.200$\pm$0.001 & 20.1$\pm$0.1  \\

NGC1140 & 1.45$\pm$0.03 & 0.125$\pm$0.001 & 18.3$\pm$0.2 & & & 1.45$\pm$0.03 & 0.125$\pm$0.001 & 18.3$\pm$0.2 \\

NGC1569 & 4.97$\pm$0.06 & 0.421$\pm$0.002 & 13.4$\pm$0.1 & & & 4.91$\pm$0.06 & 0.418$\pm$0.002 & 13.3$\pm$0.1 \\

IIZw40 & 8.07$\pm$0.15 & 0.881$\pm$0.009 & 10.0$\pm$0.1 & & & 7.91$\pm$0.15 & 0.837$\pm$0.009 & 9.74$\pm$0.08 \\

UGC4274 & 0.634$\pm$0.010 & 0.0666$\pm$0.0020 & 15.5$\pm$0.2 & & & 0.634$\pm$0.010 & 0.0683$\pm$0.0020 & 15.8$\pm$0.2 \\

IZw18 & 2.33$\pm$0.28 & 0.456$\pm$0.048 & 13.2$\pm$1.0 & & & 2.24$\pm$0.27 & 0.431$\pm$0.045 & 13.1$\pm$1.0 \\

Mrk1450 & 3.19$\pm$0.07 & 0.350$\pm$0.003 & 14.0$\pm$0.1 & & & 3.14$\pm$0.07 & 0.347$\pm$0.003 & 13.9$\pm$0.1 \\

UM461 & 7.99$\pm$1.00 & 1.25$\pm$0.05 & 12.0$\pm$0.3 & & & 7.75$\pm$0.97 & 1.16$\pm$0.04 & 11.7$\pm$0.3 \\

Mrk1499 & 1.22$\pm$0.02 & 0.147$\pm$0.002 & 18.5$\pm$0.2 & & & 1.17$\pm$0.02 & 0.123$\pm$0.002 & 18.4$\pm$0.2 \\

\enddata

\end{deluxetable}




\def\Term#1 #2 #3/{\mbox{$\,^{#1}\!#2_{#3}$ }}
\def\Termo#1 #2 #3/{\mbox{$\,^{#1}\!#2^o_{#3}$ }}
\def\sterm #1 #2 #3/{\mbox{$\,_{#3}\!^{#1}\!#2$}}


\begin{deluxetable}{cccc}	
\tabletypesize{\scriptsize}
\tablecaption{NGC 3603, 30 Doradus, N 66 Results}
\tablewidth{0pt}
\setlength{\tabcolsep}{0.07in}
\tablenum{6}
\tablehead{

  \colhead{Object}
& \colhead{Ne$^{++}$/Ne$^+$} 
& \colhead{S$^{3+}$/S$^{++}$}
& \colhead{Ne/S}

}

\startdata


NGC 3603\#3 & 0.596 & 0.0953 & 23.4 \\
NGC 3603\#4 & 0.257 & 0.0312 & 14.6 \\
NGC 3603\#5 & 0.953 & 0.101 & 13.0 \\
 & & & \\
30 Dor\#2 & 1.72 & 0.0913 & 11.9 \\
30 Dor\#3 & 3.30 & 0.197 & 11.2 \\
30 Dor\#4 & 5.98 & 0.398 & 10.8 \\
30 Dor\#5 & 1.55 & 0.103 & 11.7 \\
30 Dor\#6 & 2.06 & 0.129 & 11.3 \\
30 Dor\#7 & 4.19 & 0.274 & 11.3 \\
30 Dor\#8 & 0.791 & 0.0590 & 13.4 \\
30 Dor\#10 & 0.351 & 0.0517 & 13.8 \\
30 Dor\#11 & 2.29 & 0.122 & 11.3 \\
30 Dor\#12 & 3.27 & 0.234 & 11.4 \\
30 Dor\#13 & 2.50 & 0.147 & 10.9 \\
30 Dor\#14 & 3.28 & 0.230 & 12.1 \\
30 Dor\#15 & 3.80 & 0.219 & 11.5 \\
30 Dor\#16 & 5.64 & 0.318 & 10.8 \\
30 Dor\#17 & 0.400 & 0.0353 & 15.2 \\
 & & & \\
N 66\#1 & 4.65 & 0.285 & 8.97 \\
N 66\#2 & 3.83 & 0.294 & 9.40 \\
N 66\#5 & 3.81 & 0.240 & 9.95 \\
N 66\#6 & 1.71 & 0.250 & 10.4 \\
N 66\#7 & 5.24 & 0.252 & 9.25 \\
N 66\#8 & 2.27 & 0.194 & 10.7 \\
N 66\#9 & 2.05 & 0.164 & 11.2 \\
N 66\#10 & 2.04 & 0.167 & 11.3 \\
N 66\#11 & 2.03 & 0.160 & 10.9 \\
N 66\#12 & 5.33 & 0.447 & 10.1 \\
N 66\#13 & 3.37 & 0.225 & 9.80 \\

\enddata

\end{deluxetable}


\end{document}